\documentclass[twocolumn, twocolappendix, numberedappendix]{emulateapj}
\usepackage{color, enumerate, natbib}
\usepackage{hyperref}
\shorttitle{$\alpha$ Herculis System}
\shortauthors{Moravveji, Guinan et al.}

\newcommand{\teff}{T_{\rm eff}}
\newcommand{\dhip}{d_{\mbox{\scriptsize Hip}}}

\newcommand{\aov}{$\alpha_{\rm ov}$}

\begin{document}
\title{The Age and The Mass of The $\alpha$ Herculis Triple-Star System From A \texttt{MESA} Grid of rotating stars 
         with 1.3$\leq M/M_\odot\leq8.0$\footnote{The grid is publicly available for retrieval.}}

\author{Ehsan Moravveji\altaffilmark{1,2,3,4}}
\altaffiltext{1}{School of Astronomy, Institute for Research in Fundamental Sciences (IPM), PO Box 19395-5531, Tehran, Iran}
\altaffiltext{2}{Instituut voor Sterrenkunde, K.U. Leuven, Celestijnenlaan 200D, 3001, Leuven, Belgium}
\altaffiltext{3}{Department of Astronomy, Villanova University, 800 Lancaster Ave, Villanova PA, USA}
\altaffiltext{4}{Department of Physics, Institute for Advanced Studies in Basic Sciences (IASBS), Zanjan 45137-66731, Iran}
\email{Ehsan.Moravveji@ster.kuleuven.be}
\author{Edward F. Guinan\altaffilmark{3}}
\author{Habib Khosroshahi\altaffilmark{1}}
\author{Rick Wasatonic\altaffilmark{3}}
\shorttitle{The Age and Mass of The $\alpha$ Herculis System}
\shortauthors{Moravveji E. et. al.}
\begin{abstract}
$\alpha^1$ Her is the second closest Asymptotic Giant Branch (AGB) star to the Sun, 
and the variable luminous M5 Ib-II member of a triple stellar system containing G8 III and A9 IV-V components. 
However, the mass of this important star was previously uncertain with published values ranging from $\sim$2 - 15$M_\odot$. 
As shown by this study, its fortuitous membership in a nearby resolved triple star system, makes it possible to determine its fundamental properties including its mass and age.  
We present over twenty years of VRI photometry of $\alpha^1$ Her as well as Wing intermediate-band near-IR TiO and NIR continuum photometry.
We introduce a new photometry-based calibration technique, and extract the effective temperature and luminosity of $\alpha^1$ Her, in agreement with recent interferometric measures.
We find, $\teff=3280\pm87$ K and $\log(L/L_\odot)=3.92\pm0.14$.

With the MESA code, we calculate a dense grid of evolutionary tracks for Galactic low- to intermediate-mass (1.3 to 8 $M_\odot$) rotating stars 
from the pre-main sequence phase to the advanced AGB phase. 
We include atomic diffusion and rotation mechanisms to treat the effects of extra elemental mixing.
Based on the observed properties of the $\alpha$ Herculis stars, we constrain the age of the system to lie in the range 0.41 to 1.25 Gyr.
Thus, the mass of $\alpha^1$ Her lies in the range $2.175\leq M/M_\odot \leq 3.250$.
We compare our model-based age inference with recent tracks of the Geneva and STAREVOL codes, and show their agreement.
In the prescribed mass range for $\alpha^1$ Her, the observed $^{12}$C/$^{13}$C and $^{16}$O/$^{17}$O ratios are consistent (within 2$\sigma$) 
with the ratios predicted by the MESA, Geneva and STAREVOL codes.
\end{abstract}

\section{Introduction}\label{s-intro}
The details of how a star evolves is mainly governed by three fundamental properties it inherits from its birth place:
the chemical composition, angular momentum, and mass.
The metal content and the projected rotation velocity of a star can be measured from spectroscopy.
The mass assessments, however, are currently possible from either astereoseismic analyses of radially and/or non-radially pulsating stars, 
or more directly from the analyses of the light and radial velocity curves of double-line spectroscopic eclipsing binary systems. 
When the seismic and binary information are absent, one classically calculates a grid of stellar models, and tries to fit a range of tracks to the
observed global properties of the star, such as effective temperature, surface gravity and luminosity to estimate other physical properties like age, mass and radius.
However, this grid-based (isochronal) approach has large uncertainties associated with it \citep{basu-2012-01}; 
small uncertainties in $\teff$, Fe/H, $L/L_\odot$ translate into large uncertainties in age and mass of the stars.

In the present study, we pursue the grid-based approach, and establish the range of possible masses for the three stars (with a single age) 
in the $\alpha$ Herculis triple star system.
Our rationale is to take advantage of the fortuitous membership of the bright M5 Ib-II Asymptotic Giant Branch (AGB) star in a resolved triple star system with a 
good parallax to determine its physical properties by simultaneously fitting the observed properties.
The recent release of stellar evolutionary tracks (mainly by the Geneva group) 
provides an excellent opportunity to test the model dependence of the inferred physical properties of $\alpha^1$ Her.
For this purpose, we compare MESA with the rotating and non-rotating tracks of \cite[][hereafter E12]{ekstrom-2012-01}, 
\cite[][hereafter L12]{lagarde-2012-01} and \cite[][hereafter M12]{mowlavi-2012-01}.

The stellar mass has an additonal critical role:
The surface abundances of AGBs depend on the efficiency of the previous dredge-up episodes in addition to (extra) 
non-convective mixing mechanisms \citep[e.g.][]{karakas-2010-01, abia-2012-01}.
Different proposed extra mixing mechanisms are:
rotation and atomic diffusion \citep{maeder-2012-01}, internal gravity waves \citep{talon-2005-01}, magnetic dynamo \citep{busso-2007-01},
and thermohaline mixing after the sub-giant phase \citep{charbonnel-2007-01, cantiello-2010-01}.
The latter affects low-mass stars \citep{stancliffe-2010-01}, so, the net strength of surface enrichments depends explicitly on stellar mass.
Since $\alpha^1$ Her has had its $^{12}$C/$^{13}$C and $^{16}$O/$^{17}$O isotope ratios measured, here we provide a rare calibration 
point for these ratios at an intermediate mass and luminosity on the AGB \citep{el-eid-1994-01}.

\section{Literature Debates on the Mass of $\alpha^1$ Her}\label{s-mass-debate}
Historically, the reports on the mass of $\alpha^1$ Her from the literature are inconsistent.
They are $M = 15$ M$_\odot$ by \citet{deutsch-1956-01}, $\sim$2.0 M$_\odot$ by \cite{woolf-1963-01},  
1.7 M$_\odot$ by \citet{reimers-1977-01} and \citet{thiering-1993-01}, $\sim$5 to 7 M$_\odot$ by \citet{harris-1984-01} and \citet{el-eid-1994-01}, 
and $2.5^{+1.6}_{-1.1}\,M_\odot$ by \citet{moravveji-2011-01}.
This uncertain mass of $\alpha^1$ Her translates into its unknown evolutionary status:
assuming a high mass, it could be a red supergiant or a super-AGB star and a progenitor of iron or electron-capture core-collapse supernova 
\citep{polarends-2008-01, smartt-2009-02}, 
while in the lower mass regime ($M<$ 5 M$_\odot$), the star would be located near the upper tip of the AGB.

The membership of a bright AGB star in a nearby multiple star system is an excellent opportunity to determine the mass and evolutionary properties of an AGB star. 
Very few AGB stars have reliable ages and masses.
We previously studied the $\alpha^1$ Her light curve in \cite{moravveji-2010-01}, and extracted the dominant pulsation periods.
The purpose of this second paper is to constrain the parameter space of the global physical properties of $\alpha^1$ Her -
i.e. its mass, effective temperature and luminosity - for a subsequent asteroseismic modelling.

Firstly, we introduce the $\alpha$ Herculis system in Section \ref{s-alf-her-sys} and our photometric compilation in Section \ref{s-phot}. 
In Section \ref{s-calib} we calibrate the effective temperature based on the strength of the TiO $\lambda$7190 \AA ~absorption bands.
In Section \ref{s-lum-rad} we derive the time-variable luminosity and radius of $\alpha^1$ Her.
Based on these, we set up a dense grid of evolutionary models (Section \ref{s-alf-her-model}).
In Section \ref{s-results}, we establish the most likely range for the age of the system and masses of its individual members, and compare our findings
with those from three other codes.
As a by-product of the grid calculation, we additionally present the explicit dependence of mixing of C and O isotopic ratios on the stellar mass during the AGB phase.
We discuss our results in Section \ref{s-conclusion}.

\section{The $\alpha$ Herculis System}\label{s-alf-her-sys}

\begin{deluxetable*}{lcccccccccccc}
 \tablecaption{Compilation of the observed physical properties of stars in the $\alpha$ Her system from the literature. 
 For the results of this study refer to Table \ref{t-tlr}.\label{t-alf-sys} }
  \tabletypesize{\normalsize}
  \tablecolumns{13}
  \tablehead{  \colhead{} & \colhead{} & \multicolumn{3}{c}{$\alpha^1$ Her} & \colhead{} & 
  \multicolumn{3}{c}{$\alpha^2$ Her A} & \colhead{} & \multicolumn{3}{c}{$\alpha^2$ Her B}  \\
  \cline{3-5} \cline{7-9} \cline{11-13} \\
  \colhead{} & \colhead{} & \colhead{Value} & \colhead{Ref.} & \colhead{Note} & \colhead{} & 
  \colhead{Value} & \colhead{Ref.} & \colhead{Note} & \colhead{} & \colhead{Value} & \colhead{Ref.} & \colhead{Note} } 
  \startdata
 Spectral Class         &  & M5 Ib-II & (a,c) & (1) & & G8 III & (d) &  &  & A9 IV-V & (d) &  \\
 $\teff$ [K]              &  & 3271$\pm$46 & (g) & (2) & & 4900$\pm$150 & (d) & (3,4) &  & 7350$\pm$150 & (d) &  \\
                               &  & 3285$\pm$89 & (h) &      & &                           &      &         &  &                          &      &  \\
                               &  & 3260$\pm$40 & (i)  & (5) & &                           &      &         &  &                          &      &   \\
 $\log(L/L_\odot)$  &  & 4.25$\pm$0.30 & (h) &    & &                           &      &         &  &                          &      &   \\
                               &  & 3.68                  & (d) & (3)& & 2.10$\pm$0.04 & (d) & (3,6) &  & 1.41                  & (d) & (3) \\
 $V$ [mag]              &  & $+$3.350$\pm$0.003& (c) &      & & $+$5.6   & (a) &         &  & $+$6.6               & (f)  &   \\
 $\phi$ [mas]          &  & 33.0$\pm$0.8   & (g,i) & (5)  & &                           &      &         &  &                          &      &   \\
                              &  & 31.51$\pm$0.08  & (h) & (7) & &                           &      &         &  &                          &      &   \\
                              &  & 39.32$\pm$1.04 & (j) & (8) & &                           &      &         &  &                          &      &   \\
                              &  & 37.22$\pm$2.94 & (k) &    & &                           &      &         &  &                          &      &   \\
 $\dhip$ [pc]          &  & 110$\pm$16       & (l) &   & &                           &      &         &  &                          &      &   \\
 $\dot{M}$ [$M_\odot$ yr$^{-1}$] &  & 1.1 - 1.5 $\times 10^{-7}$ & (d,e) &  & &                           &      &         &  &                          &      &   
  \enddata
  \tablerefs{
  (a) \cite{deutsch-1956-01}, (b) \cite{keenan-1989-01}, (c) This study, (d) \cite{thiering-1993-01},
  (e) \cite{reimers-1977-01}, (f) \cite{woolf-1963-01}, (g) \cite{dyck-1996-01}, (h) \cite{perrin-2004-01},
  (i) \cite{benson-1991-01}, (j) \cite{weiner-2003-01}, (k) \cite{richichi-2002-01} (l) \cite{vanleeuwen-2007-01}. 
  }
  \tablecomments{
  (1) From Coud\'{e} Spectroscopy of \cite{deutsch-1956-01}, 
  (2) See Figure \ref{f-tlr}, Table \ref{t-tlr} and Section \ref{s-lum-rad}.
  (3) We take the mean $L/L_\odot=10.5\pm4.5$ from \cite{thiering-1993-01}, and correct for the underestimated distance (70 pc). 
  (4) An error of $\pm$150 K in temperature for $\alpha^2$ Her A is assumed. 
  (5) K-band interferometry ($\lambda_0=2.2\mu$m, $\Delta\lambda=0.4\mu$m), 
  (6) An assumed error of 0.1 mag in $m_V$, translates to an error of $\pm12 L_\odot$, 
  (7) Corrected to the revised Hipparcos distance, K-band interferometry, 
  (8) Interferometry in mid-infrared ($\lambda_0=$9.5 to 11.5$\mu$m). 
  The mid-infrared diameter of AGBs is up to $\sim$30\% larger than their corresponding near-infrared diameter.
  }
  \end{deluxetable*}

$\alpha$ Herculis is an extensively studied system, and is composed of three stars.
Based on the literature, several properties of the system is known, up to varying accuracies 
\citep[e.g.][]{deutsch-1956-01, reimers-1977-01,thiering-1993-01}.
\citet{mcalister-1989-01} speculate the presence of the fourth or even fifth members. 

The primary $\alpha^1$ Her (Rasalgethi, HD 156014, $V=3.350\pm0.003$ mag, $K=-3.511\pm0.150$ mag) is an M5 Ib-II \citep{deutsch-1956-01, keenan-1989-01} 
semi-regularly pulsating bright giant.
According to the \citet{morgan-1973-01} classification, the spectra of $\alpha^1$ Her is a standard for its subclass.
The secondary, $\alpha^2$ Her  ~(HD 156015, $V=5.39$) is a spectroscopic binary itself \citep{thiering-1993-01}, consisting of a G5 III giant 
(hereafter $\alpha^2$ Her  A) and an A9 IV-V dwarf (hereafter $\alpha^2$ Her  B). 
The primary and the secondary are 4.7 arcsec distant \citep{jeffers-1978-01}, so our photometry of the system (Section \ref{s-phot}) includes the flux from three members. 
Where necessary, we have replaced the first Hipparcos parallax of \citet[][$\pi=8.53\pm2.80$ mas]{perryman-1997-01} with the revised value of \cite{vanleeuwen-2007-01};
thus, the distance to the system is
%

\begin{equation}\label{e-alf-her-hip}
\pi = 9.07\pm1.32 \mbox{ mas} \Rightarrow \dhip = 110\pm16 \mbox{ pc.}
\end{equation}
With the parallax $\pi$ and the disk angular diameter $\phi$ expressed in milliarcsec (mas) the limb-darkened radius of the star can be assessed by

\begin{equation}\label{e-phi-R}
R=107.55\frac{\phi}{\pi}\, R_\odot
\end{equation}

The angular diameter of $\alpha^1$ Her is already measured by different interferometry groups.
Figure 1 in \citet{perrin-2004-01} addresses the strong wavelength dependence of the angular diameter measurement, from near- to mid-IR \citep{weiner-2003-01}.
The average of the limb-darkened K-band interferometry of \citet{perrin-2004-01} - from 1996 to 1997 - is $\phi=31.51\pm0.08$ mas.
Similar assessment of \citet{richichi-2002-01} yields $\phi=37.22\pm2.94$ mas, and \cite{weiner-2003-01} give $\phi=39.32\pm1.04$.
\cite{benson-1991-01} measurement gives almost consistent angular diameter $\phi=33.0\pm0.8$ mas \citep[similar to][]{dyck-1996-01}.
Adopting the angular diameter measure of \cite{perrin-2004-01}, the near-IR limb-darkened interferometric radius of $\alpha^1$ Her is 
\begin{equation}\label{e-alf-her-R}
R_{\rm inter}=400\pm 61\, R_\odot.
\end{equation}
The large error in the radius is dominated by the parallax uncertainty.
Also, as shown in this study, the diameter of $\alpha^1$ Her varies by up to $\sim$14\%.
Additionally, \cite{perrin-2004-01} find $\teff= 3285\pm89$ K.
Correcting for the revised 2007 Hipparcos parallax, we estimate the luminosity of $\alpha^1$ Her as $\log(L/L_\odot)=4.25\pm0.30$.\footnote{
Using Eq. \ref{e-alf-her-R} and the Stefan-Boltzmann law, $\log(L_{\rm new}/L_{\rm old})=2\log(R_{\rm new}/R_{\rm old})=2\log(d_{\rm new}/d_{\rm old})\approx-0.05$ dex.}

Spectroscopy of \citet{deutsch-1956-01} shows that the $\alpha$ Her system is enshrouded in an envelope of dust.
This was later  confirmed by the observations of \citet{thiering-1993-01} that the extent of the envelope is larger 
than the semi-major axis of the visual binary orbit. 
Recently, interferometric observations of \citet{tatebe-2007-01} at $\lambda_0=11.5\mu$m over the period 1989-2004 
show that $\alpha^1$ Her has experienced a major outburst during 1990 in which $\sim10^{-6}M_\odot$ has been ejected into the ISM, with an 
approximate ejecta speed of 75 km s$^{-1}$. 
The same study finds that the shell has a temperature of 518 K, and an inner and outer  angular radius of 250 and 350 mas, respectively. 
The mass loss rate of $\alpha^1$ Her is $\dot{m}=1.1-1.5\times10^{-7}M_\odot$ yr$^{-1}$ 
\citep{reimers-1977-01, thiering-1993-01}.  

Table \ref{t-alf-sys} summarizes a collection of physical parameters of $\alpha$ Herculis stars.
They are relevant to our study, and are collected from the literature.
Where necessary, we have corrected  the first Hipparcos parallax of \citet[][$\pi=8.53\pm2.80$]{perryman-1997-01} with the revised value of \citet{vanleeuwen-2007-01}.

For $\alpha^1$ Her, the spectroscopic measurements of the surface yields of CNO-processed elements date back to the studies of 
\cite{thompson-1974-01} and \cite{harris-1984-01}.
These observational evidences - when directly compared to the theoretical model yields - help gauging the role of different flavours of composition mixing 
\citep[e.g. ][]{el-eid-1994-01,cantiello-2010-01}.
We adopt $^{12}$C/$^{13}$C = $17\pm4$, $^{16}$O/$^{17}$O = $180_{-50}^{+70}$ and $^{16}$O/$^{18}$O = $550_{-175}^{+225}$ from \cite{harris-1984-01}.

\section{Multi-Color Photometric Observations}\label{s-phot}
The photometric data for the $\alpha$ Her system was obtained for more than two decades, and the data are compilations of observations from two different sites.
The photometry has been conducted with the broad-band Johnson V-filters and three intermediate-band Wing ABC 
filters \citep{wing-1992-01} in Villanova University (VU) by two 20- and 28-cm Schmidt Cassegrain telescopes.
The broad-band Johnson VRI photometry was obtained using the Fairborn-10 (T2) Automatic Photometric Telescope
\citep[APT,][]{henry-1995-01} at Tennessee State University (TSU). 
The starting and ending observation dates for each dataset are different. 
The sampling is not regular and depends on the visibility of the star and the weather condition. 
The TSU observations commence in March 1986, and were discontinued in June 2001. 
The TSU photometry are first published in \cite{percy-2001-01}. 
For the details of our dataset and the observations time baseline see Table \ref{t-obs}. 
In this table, the first column gives the designations for the filters.
The second column gives the central passband wavelength $\lambda_0$ accompanied with its corresponding full width at half maximum (FWHM).
The third column shows the observation site.
The fourth column gives the maximum and minimum of the magnitude in the corresponding filter for the entire observations due to stellar variability.
Note that the variability at shorter wavelengths has larger amplitudes.
The fifth column is the standard error.
The sixth column gives the start and end dates for the observations at the corresponding site, $T_{\mbox{\scriptsize start}}$ and $T_{\mbox{\scriptsize end}}$, respectively.
They are expressed in modified Julian date, MJD $=$ HJD $-$ 2\,400\,000.
The seventh column gives the Rayleigh limit $1/\Delta T$ in units of $10^{-4}$d$^{-1}$, where $T$ is the observation time baseline in days.
The last column gives the number of observations taken over the prescribed duration per each site.
This photometry includes all three components in the measure but the brightness is dominated by the bright, luminous M5 Ib-II star.

\placetable{Please place Table 2 on top of page 3}
	\newcommand{\ts}{$T_{\mbox{\scriptsize start}}$}
	\newcommand{\te}{$T_{\mbox{\scriptsize end}}$}
    \begin{table*}[t!]
    \normalsize
    \caption{Photometry of $\alpha$ Her system in different filters in increasing central wavelength $\lambda_0$ order.} 
    \label{t-obs}
    \begin{tabular}{lccccccc}
    \hline 
     Filter & wavelength $\lambda_0$\tablenotemark{a} 		& Observation & Max; Min  & Std. Error $\sigma$ & \ts; \te   	      & $1/\Delta T$         & $N$ \\
              & (FWHM) [\AA]                                  & site\tablenotemark{b}     & [mag]	& [mag]	& [MJD]\tablenotemark{c} & [$10^{-4}$ d$^{-1}$] &  	\\
     \hline 
     Johnson V\tablenotemark{d} & 5500 (700) & VU & $+2.768$;  $+3.624$ & 0.003  & 49043; 55076 & 1.657 & 728 \\
     Johnson V\tablenotemark{d} & 5500 (700) & TSU& $+2.922$;  $+3.792$ & 0.004 & 46510; 52089 & 1.792 & 1766 \\
     Johnson R                  & 6400 (1400) & TSU& $-2.993$;  $-2.437$ & 0.002 & 46510; 52089 & 1.792 & 1757 \\
     Wing-A (TiO)               & 7190 (110) & VU & $+0.093$;  $+0.817$ & 0.005 & 50489; 55076 & 1.657 & 547 \\
     Wing-B\tablenotemark{e}         & 7540 (110) & VU & $-1.519$;  $-1.012$ & 0.003 & 50489; 55076 & 1.657 & 547 \\
     Johnson I                  & 8800 (1500) & TSU& $-3.748$;  $-3.455$ & 0.001 & 46510; 51993 & 1.823 & 1697 \\
     Wing-C\tablenotemark{f}           & 10400 (420)& VU & $-1.707$;  $-1.449$ & 0.002 & 50489; 55076 & 1.657 & 547 \\
    \tableline
    \end{tabular}
    \end{table*}

Both the VU and TSU observations were conducted \textit{differentially} with respect to comparison stars.
The offset was removed by finding the shift between the two datasets that minimizes the standard deviation of the combined dataset during the observed overlapping runs.
The compiled light curves in Johnson V and Wing ABC filters are shown in the Appendix (Figure \ref{f-obs-all}). 
Note that, from shorter to longer wavelengths,  
the star appears brighter \citep[see][and fourth column in Table \ref{t-obs}]{benson-1991-01}, and the amplitude of light variability decreases.
This is similar to the pulsation behaviour of Mira-type stars \citep[e.g.][]{lockwood-1971-01}.

\begin{figure}[b]
\centering{
\includegraphics[width=\columnwidth]{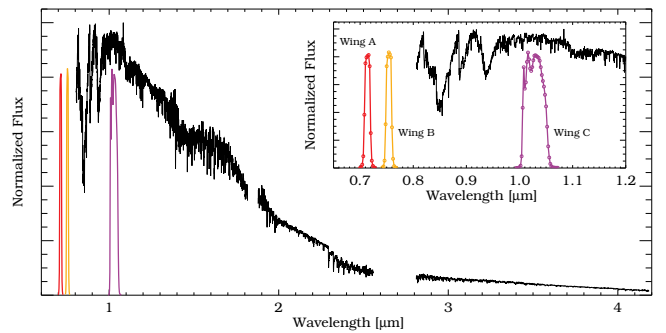}
\caption{Near-IR spectra of $\alpha$ Her taken from \cite{rayner-2009-01}. 
The transmission functions for Wing ABC filters are plotted. 
The Wing C filter sees through the peak of the continuum.}\label{f-spec}
}
\end{figure}

Figure \ref{f-spec} presents the low-resolution spectra of $\alpha$ Her published by \cite{rayner-2009-01}.
It covers the wavelength range of 0.8 to 4.2 micron.
The Wing ABC transmission functions $S_\lambda$ are also plotted.
These three filters were selected by Robert Wing (private communication) for measuring temperatures and luminosities of evolved M-type stars.
The TiO ($\gamma$, 0, 0) $\lambda$7190 absorption band strength is very sensitive to the temperature for evolved M-stars.
The A-filter with $\lambda_0$=7190 \AA ~is centered on the TiO band, and serves as a reference measure of the TiO band strength.
The B-filter at $\lambda_0$=7540 \AA ~is located essentially on the continuum region. 
The measured flux in C-filter at $\lambda_0$=10400 \AA ~can be corrected to give the bolometric magnitude $m_{\rm bol}$ and luminosity $L$ (Section \ref{s-lum-rad}). 
In the Appendix \ref{app-abm}, we show that this bolometric correction to Wing C filter is in fact BC$_{\rm C}=1.735\pm0.030$ mag.
The main conclusion from Figure \ref{f-spec} is that the Wing C filter does not suffer from strong absorption bands, and measures the peak of continuum of an M5 AGB.
Compared to the Johnson V-bandpass, it also suffers less from strong TiO absorption bands. 


\section{Effective Temperature Calibration}\label{s-calib}

\subsection{Input standard stars}\label{ss-standards}
\citet{levesque-2005-01} tabulate their findings on spectrophotometric temperature calibration for 74 galactic red supergiants based on $(V-K)_0$
color combined with synthetic MARCS stellar atmosphere models. 
For bright M5 AGBs, their Table 5 gives $T_{\mbox{\scriptsize eff-M5 I}}=3450$ K, which does not agree with 
$\teff$ derived from interferometry (Table \ref{t-alf-sys}).
However, $\alpha^1$ Her is the only M5 star in their list, thus the derived $\teff$ and V-filter bolometric correction (BC$_{\rm C}$)  
for such late type stars might be subject to a bias. 
Therefore, the M5 entry in Table 5 of \cite{levesque-2005-01} is considered unreliable.
Therefore, we do not rely on this M5 entry in \cite{levesque-2005-01}.
Our independent calibration yields $\sim200$ K cooler $\teff$ for M5 AGBs.

Eighteen standard stars were selected from \cite{wing-1978-01}, and were observed at the VU site through Wing ABC filters, repeatedly. 
This helps defining two color indices - $\gamma_1$ and $\gamma_2$ - for each of these stars. 
Following \cite{wing-1992-01}, the $B-C$ color index 
\begin{equation}
\gamma_1 = (B-C), \label{e-gamma1}
\end{equation}
in M-type giant stars is sensitive to temperature variations, since it tracks the slope of the tail of Planck distribution.
Hence, it can be calibrated to yield the effective temperature of such late type stars. 
Yet, some absorption bands may interfere.
Thus, the other color index which is called the TiO index $\gamma_2$ and defined as
\begin{equation}
\gamma_2 = A - B -0.13\gamma_1. \label{e-gamma2}
\end{equation}
is less affected by TiO absorption lines, and tracks the changes in the temperature better \citep{wing-1992-01}. 
The complete list of Wing standard stars, along with their measured mean $\bar{\gamma}_2$  is presented in Table \ref{t-standard-stars}, where
the first column is the identification number, and the second column gives the stars' HR designation.
The third column gives the spectral classification taken from \cite{wing-1978-01}.
The fourth column gives the effective temperatures taken from \cite{levesque-2005-01}.
The last column is the average $\bar{\gamma}_2$ for each of the program stars.
Since $\gamma_2$ could be time variable for the standard stars as for $\alpha^1$ Her, we average over their $\gamma_2$ values during our long-term monitoring.
The list of program stars is sorted in the decreasing $\teff$.

\begin{figure*}[t!]
\centering{
\includegraphics[width=\textwidth]{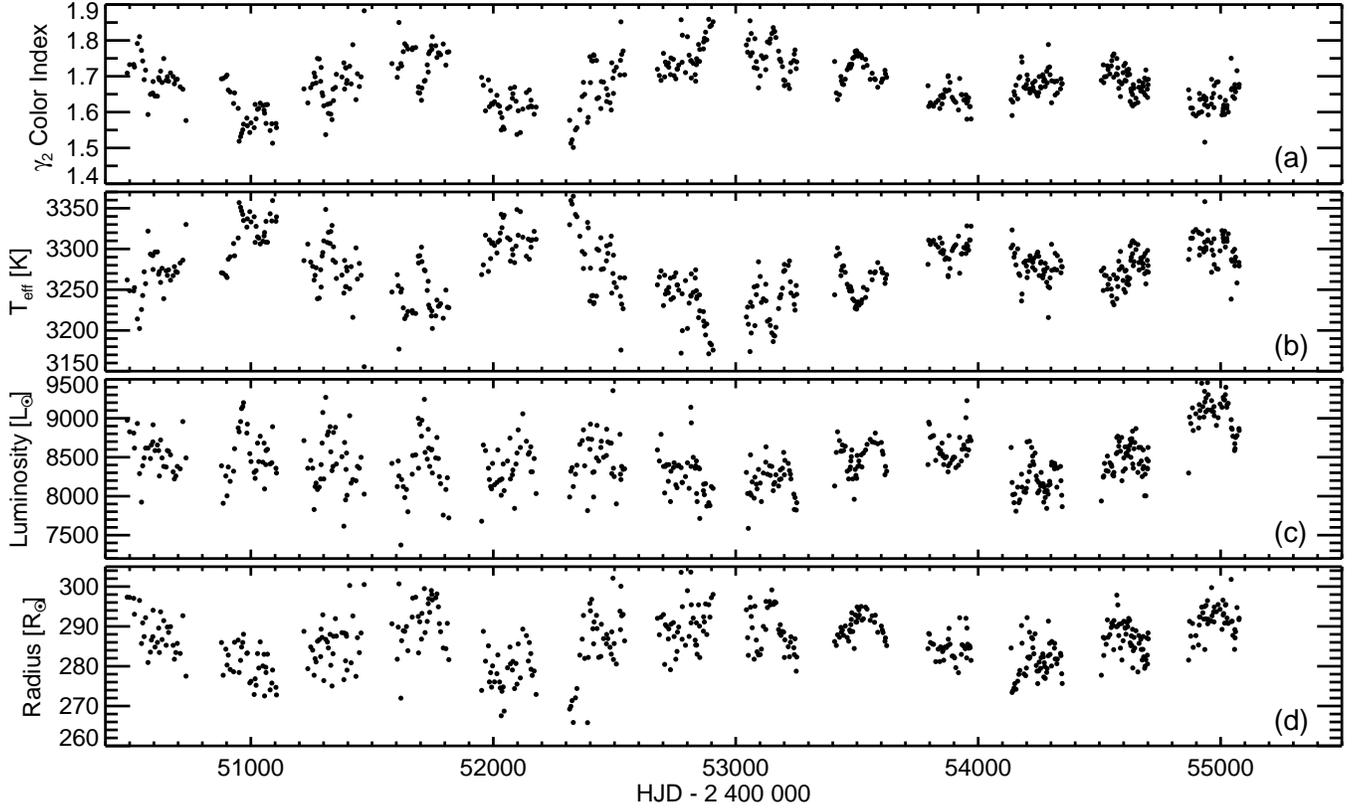}
\caption{Temporal variation in the $\gamma_2$ color index (a), effective temperature (b), luminosity (c) and radius (d). 
 The variability with the period of LSP ($\sim1400$ d) in all physical parameters is evident. 
 See also Eqs. \ref{e-gamma2}, \ref{e-bol-corr-c}, \ref{e-bol-abs-mag} and \ref{e-bol-lum}.
 The full light curves are published in the Appendix Figure \ref{f-obs-all}.}
\label{f-tlr} }
\end{figure*}

Figure \ref{f-tlr}a shows the time variability of $\gamma_2$. 
Due to the observed Long Secondary Period \citep[LSP, ][]{kiss-2006-01,percy-2001-01} of $\alpha^1$ Her, $\gamma_2$ varies with the period of $\sim1400$ days; 
this can serve as an evidence for the pulsation origin of the LSP. 

\begin{table}[t]
\begin{center}
\caption{The list of standard stars from \cite{wing-1978-01}.} \label{t-standard-stars}
\normalsize
\begin{tabular}{l|cccc}
\hline \hline
ID & HR Number & Spectral\tablenotemark{a} &  $\teff$\tablenotemark{b} & $\bar{\gamma}_2$\tablenotemark{c} \\ 
    &                    & Class                                 &  [K]                                   & [mag]            \\ 
\hline 
1  &  6705  &            K5.0  &         3940   &          0.202 \\
2  &    248  &            K5.4  &         3920   &          0.222 \\ 
3  &    337  &            M0.5  &         3934   &          0.349 \\
4  &  8284  &            M1.0  &         3817   &          0.400 \\
5  &      48  &            M1.5  &         3778   &          0.474 \\
6  &      45  &            M2.0  &         3736   &          0.560 \\
7  &    750  &            M2.5  &         3690   &          0.627 \\
8  &  9064  &            M3.0  &         3641   &          0.759 \\
9  &  9089  &            M3.4  &         3599   &          0.892 \\
10 &    211 &             M4.1 &          3522  &           1.026 \\
11 &  4483 &             M4.5 &          3475  &           1.250 \\
12 &  4909 &             M5.1 &          3401  &           1.424 \\
13 &    587 &             M5.1 &          3401  &           1.508 \\
14 &   5512 &            M5.5 &          3348  &           1.576 \\
15 &   4267 &            M5.9 &          3294  &           1.725 \\
16 &   7941 &            M5.9 &          3294  &           1.618 \\
17 &   6146 &            M6.6 &          3194  &           1.694 \\
18 &   3639 &            M7.1 &          3118  &           1.941 \\
\hline 
\end{tabular}
\end{center}
 (a) The spectral classes are assigned by \cite{wing-1978-01},
 (b) $\teff$ is taken from \cite{levesque-2005-01} for mid-K to mid-M giants and supergiants, 
 (c) $\bar{\gamma}_2$ is measured at VU.
\end{table}

\subsection{Calibrating $\teff$ versus $\gamma_2$ Color Index}\label{ss-fitting}
\begin{figure}
\includegraphics[width=\columnwidth]{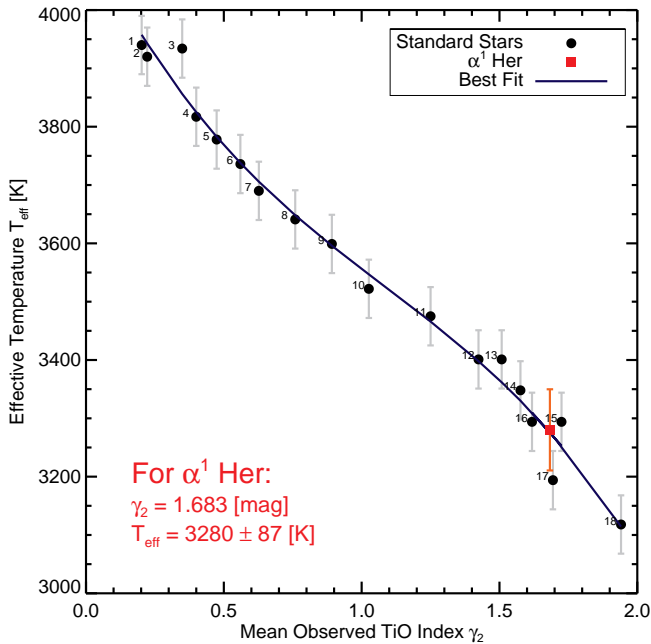}
\caption{$T_{\mbox{\scriptsize eff}}$ calibration with eighteen mid-K to late-M standard stars selected from \cite{wing-1978-01}. 
Filled circles designate the stars from Table \ref{t-standard-stars} marked with their associated ID. 
The solid curve is the polynomial fit from Eq. \ref{e-teff-calib}.
The red square represents $\alpha^1$ Her.
}
\label{f-teff-tio}
\end{figure}

To arrive at a reasonable calibration for $\teff$ versus TiO index $\gamma_2$, we use a 
Levenberg-Marquardt least-squares \citep{markwardt-2009-01} $3^{\rm rd}$ order polynomial fit 
to the entries in Table \ref{t-standard-stars}, and derive the best fit coefficients. 
The reduced chi-square goodness of fit is $\chi^2_{\rm red}$ = 1.02.
Very similarly, $\gamma_1$ could also be used, but we prefer $\gamma_2$
for its higher sensitivity to temperature changes due to TiO absorptions. 
Therefore, we end up with the following relation
\begin{equation}\label{e-teff-calib} 
\teff = 4129 (\pm 5) - 952 (\pm 20) \,\gamma_2 + 547 (\pm 22) \,\gamma_2^2 - 168 (\pm 7)\, \gamma_2^3.
\end{equation}
Compare this with last equations in \citet{levesque-2005-01}. 
The numbers in the parentheses are the 1-$\sigma$ uncertainties for each of the fitting coefficients.
The resulting fit is shown as a solid line in Figure \ref{f-teff-tio}. 
The average of the TiO index for $\alpha^1$ Her is $\gamma_2 = 1.683\pm0.003$ mag. 
Consequently, the average effective temperature after substituting mean $\gamma_2$ into Eq. \ref{e-teff-calib} is 
$\teff=3280\pm87$ K.
The uncertainties are evaluated by a Monte Carlo simulation. 
For the calibration stars in Table \ref{t-standard-stars}, the standard deviation in $\teff$ is $\pm$31 K, 
and agrees with the assumed error estimates of \cite{levesque-2005-01}.
The agreement between our indirect derivation of $\teff$ and direct interferometric measures (Table \ref{t-alf-sys}) is convincing.

Therefore, we utilize this calibration for determining $\teff$, and calculate the temperature for individual values of $\gamma_2$ at any
given epoch for $\alpha^1$ Her.
The observed maximum and minimum values of $\gamma_2$ are 1.500 and 1.881 mag, respectively.
Consequently, the upper and lower limits of the effective temperature of $\alpha^1$ Her are $\teff = $3365 K and 3155 K, respectively;
they correspond to inferred spectral types of $\sim$M5 and M6, respectively.
This temperature variation indicated by the variability in the $\gamma_2$-index, can be induced by pulsations.
This can be seen from the inferred variations of the star's radius and luminosity (Figure \ref{f-tlr}).
However, smaller non-periodic contributions to this variability could arise from the growth and decline of starspots from the changes in $\teff$
produced by the presence of large convective cells in the star's atmosphere \citep{stothers-2010-01}.
 

\section{Time Variability of Luminosity and Radius}\label{s-lum-rad}
Once the change in the color temperature is accounted for, the calculation of luminosity and radius is straightforward from the Stefan-Boltzmann law
$L/L_\odot = (R/R_\odot)^2\,(T/T_\odot)^4$. 
\cite{lattanzio-2004-01} and \cite{unno-1989-book} argue that this relation yields reliable results for the AGBs.
With a different calibration, the same law is employed in 
interferometric observations of nearby Miras and supergiants as a means of direct measurement of their luminosity and radius 
\citep[e.g.][]{weiner-2003-01, perrin-2004-01, lacour-2009-01}.
Therefore, 
\begin{eqnarray} 
&&m_{\mbox{\scriptsize bol}} = C+\mbox{BC}_{\rm C} \label{e-bol-corr-c} \\
&&M_{\mbox{\scriptsize bol}} = m_{\mbox{\scriptsize bol}} - 5.207 \label{e-bol-abs-mag} \\
&&L/L_\odot=10^{(4.75-M_{\mbox{\scriptsize bol}})/2.5}, \label{e-bol-lum}
\end{eqnarray}
%
%
\noindent $M_{V\odot} = $4.75 is the absolute magnitude of the Sun \citep{allen-1976-01}, 
5.207 is the distance modulus to the $\alpha$ Her system from Hipparcos (Eq. \ref{e-alf-her-hip}), and BC$_{\rm C}=1.735\pm0.030$ mag is 
the bolometric correction to Wing C filter (Appendix \ref{app-abm}).
The ISM absorption in the Wing C-filter along the $\alpha$ Herculis line-of-sight was deemed insignificant.
\cite{dyck-1996-01} also assume zero extinction in near-IR towards the $\alpha$ Her system\footnote{
But, we are aware that this can impose a bias in the inferred bolometric luminosity.}.
From Table \ref{t-obs} and Appendix \ref{app-abm} the bolometric apparent magnitude has a net error of $\Delta m_{\rm bol}\approx0.036+0.002=0.038$; 
therefore, a rough estimate of the uncertainty in the absolute bolometric magnitude is $\Delta M_{\rm bol}=\Delta m_{\rm bol}+
\frac{5}{\ln10}\frac{\Delta d_{\rm Hip}}{d_{\rm Hip}}=0.35$ mag, and that of luminosity is $\Delta \log(L/L_\odot)=\Delta M_{\rm bol}/2.5=0.14$ dex.
The radius variation from Stefan-Boltzman law is
\begin{equation}\label{e-blackbody} 
R/R_\odot=(L/L_\odot)^{1/2}(\teff /5779)^{-2}. 
\end{equation}
and the relative error in radius is approximately 21\%.

Figures \ref{f-tlr}.a to \ref{f-tlr}.d show how the time dependence of the TiO index $\gamma_2$ is translated to temporal
variations in physical quantities of the star $\teff$, $L/L_\odot$, and $R/R_\odot$ with the period of LSP.
When the star is hotter, it is more luminous and smaller.
Apparently, stellar pulsation is the most likely mechanism to explain the observed simultaneous variability in temperature,
luminosity and radius of the star \citep{wood-2004-01,nicholls-2009-01}.
This is a subject of a forthcoming paper.

    \begin{table}[t]
    \begin{center}
    \caption{Extrema measures of $T_{\mbox{\scriptsize eff}}$ ($^\circ$K), $L/L_\odot$, and $R/R_\odot$ for $\alpha^1$ Her
    (see Eqs. \ref{e-teff-calib} to \ref{e-blackbody}). } \label{t-tlr}
    \normalsize
    \begin{tabular}{c|cccc}
    \hline  \hline 
     & Mean & Min  & Max & $\delta$\\
    \hline
    $\teff\;(^\circ $K) & $3280\pm87$ & 3155  & 3365 & 6.4\% \\
    $log(L/L_\odot)$  & $3.92\pm0.14$  & 3.86 & 3.97 & 25.8\% \\
    $R/R_\odot$  & $284\pm60$   & 264   & 303   & 13.9\% \\
    \hline 
    \end{tabular}
    \end{center}
    \end{table}

Table \ref{t-tlr} summarizes the minimum, maximum, and average values for these calculated quantities; 
in the last column, $\delta=|$Max$-$Min$|/$Mean is the relative change in any 
quantity during our observations. 
Our derived luminosity is close to the lower limit of \cite{perrin-2004-01} (see Table \ref{t-alf-sys}), and agrees within the error bars.
The angular diameter of the star based on Eqs. \ref{e-alf-her-hip}, \ref{e-phi-R} and \ref{e-blackbody} is 23.95$\pm$5.03 mas;
this is 24\% less than the K-band angular diameter measure of \cite{perrin-2004-01} (see Table \ref{t-alf-sys}).
Despite the significant disagreement between our inferred angular diameter of $\alpha^1$ Her and that of literature (Table \ref{t-alf-sys}),
we show in Section \ref{ss-masses} that our radius assessment has a better agreement with our evolutionary models.

\section{Modeling the $\alpha$ Herculis Stars}\label{s-alf-her-model}
We assume that the three stars of the $\alpha$ Herculis system are coeval. 
We adopt the Solar chemical composition of \cite{asplund-2009-01}, i.e. $(X, Y, Z)=(0.720, 0.266, 0.014)$ \citep[see Table 1 in][and details therein]{ekstrom-2012-01}.
This choice is supported by the spectroscopy of \cite{hoflich-1986-01}.
The differences among spectroscopic classes of $\alpha$ Herculis stars (Table \ref{t-alf-sys}) imply that their initial ZAMS masses are different.
Based on the measurements collected from literature and within their corresponding uncertainties (Table \ref{t-alf-sys}), 
we model the three stars using the state-of-the-art stellar structure and evolution code 
MESA\footnote{MESA is an open-source code accessible from \href{http://mesa.sourceforge.net/index.html}{http://mesa.sourceforge.net}.
The Fortran 90 inlists and modules are also available via \href{http://mesastar.org/}{http://mesastar.org/}. 
The calculated tracks can be retrieved by directly contacting the corresponding author.} \citep[v.4589,][]{paxton-2011-01, paxton-2013-01}.
Our choice of abundances and other parameters allow us compare our models with recent results of \cite{lagarde-2012-01}, \cite{ekstrom-2012-01} and \cite{mowlavi-2012-01}.
Below, we discuss the physical ingredients of our MESA grid.
\subsection{Rotational Mixing}\label{ss-rot}
The initial equatorial rotation rate of $\alpha$ Herculis stars are unknown a priori. 
Therefore, we set up a dense grid (in initial mass) of evolutionary models that take into account the shellular rotation \citep{heger-2000-01,heger-2005-01}. 
Yet, the choices for the initial rotation rates could be various \citep[compare, e.g.][]{tassoul-2000-book, charbonnel-2010-01, cantiello-2010-01}.
Similar to \cite{lagarde-2012-01} we adopt $\eta_{\rm rot}=\Omega_{\rm eq}/\Omega_{\rm cri}=0.45$ on the ZAMS, where the critical angular velocity is
$\Omega_{\rm cri}=(8GM/27R_{\rm eq}^3)^{1/2}$.
$R_{\rm eq}$ is the equatorial radius calculated for a non-rotating case, and $M$ is the stellar mass.

\subsection{Convective, Overshoot and Thermohaline Mixing}\label{ss-mixing}
The mixing processes near the stellar core will have an appreciable effect on the duration and width of the main-sequence (MS) phase in the HR diagram \citep{maeder-2009-book}.
For our case, the mixing parameters of $\alpha^2$ Her B critically influenced the age of this star. 
The convective mixing is treated using the Mixing Length Theory (MLT) of \cite{bohm-vitense-1958-01}, with 
$\alpha_{\rm MLT}=1.6$. 
Boundaries of convective zone(s) are located where $\nabla_{\mbox{\scriptsize rad}}=\nabla_{\mbox{\scriptsize ad}}$.
The overshooting beyond the boundaries of convective zones are included with the extent of the overshoot zone a multiple of the local pressure scale height, say
$d_{\rm ov}=$\aov$H_p$ with \aov=0.10.

Thermohaline mixing has been recently discussed as a source of extra mixing in models of red giant branch (RGB) stars. 
MESA uses the formulation by \cite{kippenhahn-1980-01} and \cite{traxler-2011-01}.
For applications, see \cite{cantiello-2010-01} and \cite{charbonnel-2010-01}.
2D and 3D hydrodynamic simulations of this double-diffusive instability indicate a very slow mixing process acting in low-mass stars 
\citep{denissenkov-2010-01, denissenkov-2011-01, traxler-2011-01}.
Thus, while this might have an impact on observable surface abundances, the effect on the internal thermal structure (hence
luminosity and stellar age) is predicted to be negligible \citep{denissenkov-2008-01}.
In this study, we ignore the thermohaline mixing.
Furthermore, we also ignore the extra mixing induced by magnetic fields, but we do include the radiative levitation
based on \cite{thoul-1994-01} and \cite{morel-2002-01}. 

\subsection{Mass Loss}\label{ss-mdot}
Dust-driven mass loss from highly luminous cool stars 
depends sensitively on the mass, radius, luminosity, and metallicity of the star \citep[e.g. ][]{vanloon-2006-01}. 
We employ the \cite{reimers-1977-01} criteria for RGB mass loss, and the  
prescription by \cite{blocker-1995-01} on the AGB phase
\begin{equation}\label{e-mdot}
\begin{array}{ll}
\dot{m} = 1.4\times10^{-13} \, \eta_{\rm RGB} \, (L/gR);                     & \eta_{\rm RGB} = 0.5, \\
\dot{m} = 4.83\times10^{-9} \, \eta_{\rm AGB} \, (L^{2.7} / M^{2.1});  & \eta_{\rm AGB} = 0.1. 
\end{array}
\end{equation}
with $L$ and $M$ expressed in solar units.
The transition between the two prescriptions is made when the He mass fraction in the core is less than $10^{-3}$.
The rotationally enhanced mass loss rate is employed, similar to \cite{maeder-2001-01}.

\section{Results}\label{s-results}

\subsection{The Composite HR Diagram of $\alpha$ Herculis System}\label{ss-hrd}
We calculate a dense grid of evolutionary models comprising of 55 tracks. 
The employed mass range $M$ and stepsize in unit of $\Delta_M$ in $M_\odot$ is
\begin{equation}\label{e-grid-mass}
\begin{array}{ll}
M:\, 1.300 \cdots 1.500, & \Delta_M = 0.100 \\
M:\, 1.600 \cdots 2.300, & \Delta_M = 0.025 \\
M:\, 2.500 \cdots 8.000, & \Delta_M = 0.250
\end{array}
\end{equation}

\begin{figure*}
\centering{
\includegraphics[width=\textwidth]{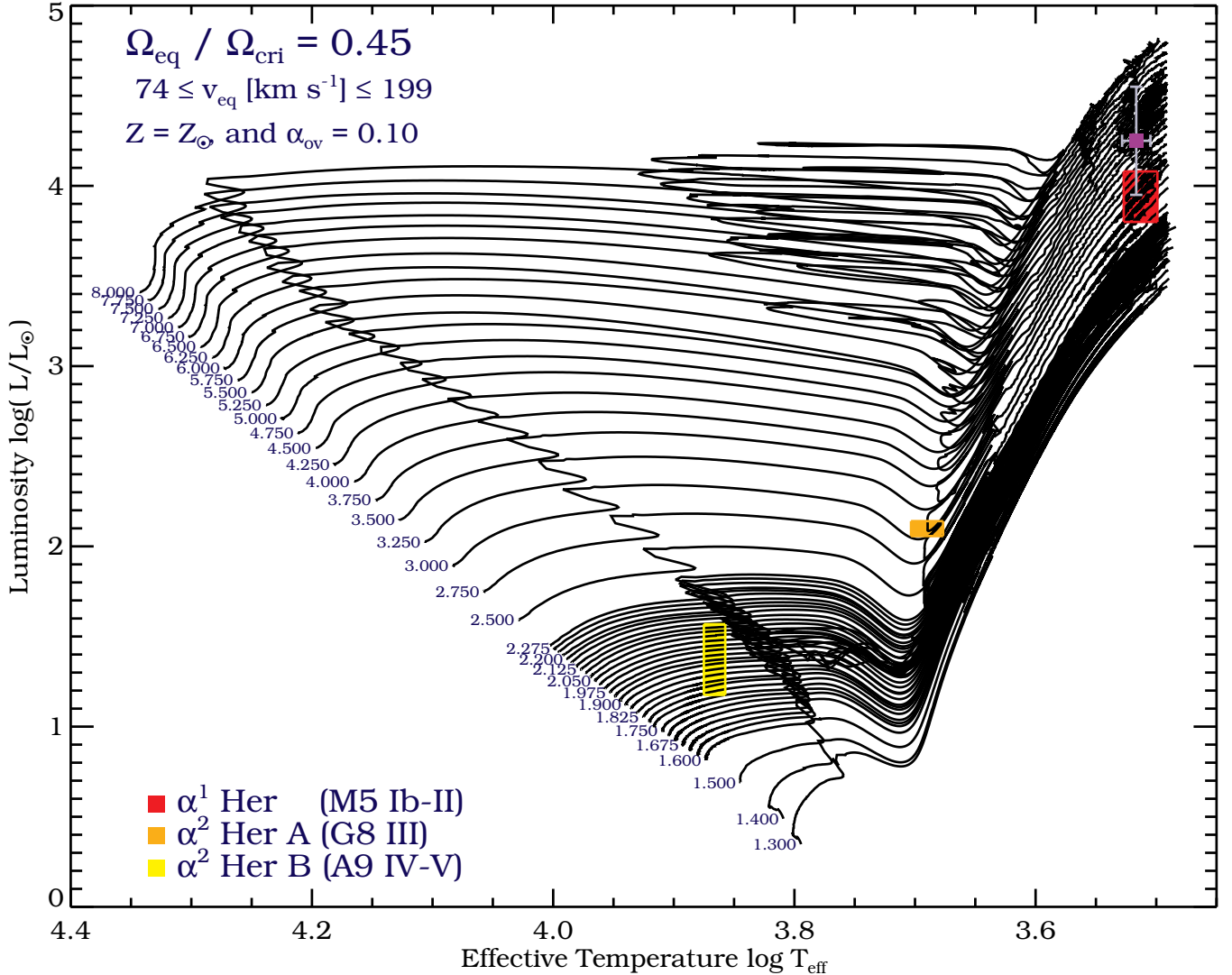}
\caption{MESA grid of rotating low- to intermediate-mass stars from the MS phase up to the AGB phase (CHeD and $\teff\leq 3100$ K)
at solar metallicity.
The initial rotation rate is $\eta_{\rm rot}=0.45$
The position of the three members of the system are highlighted based on Tables \ref{t-alf-sys} and \ref{t-tlr}.}
\label{f-hrd-all}}
\end{figure*}

For every track, the evolution calculation is stopped after the core helium depletion (hereafter CHeD), when $\teff$ drops below 3100 K.
At the end, the grid consists of more than 353\,000 rows of evolutionary information, such as $\teff$, $L$, etc. 
The synthetic absolute bolometric magnitude $M_{\rm bol}$, V-band bolometric correction B.C., and the standard Johnson-Cousins $UBV\,RI\,JHKLL'M$ color indices
are calculated based on \cite{lejeune-1998-01}.
The grid is sketched in Figure \ref{f-hrd-all}.
The positions of the three $\alpha$ Herculis members within their 1$\sigma$ boxes of uncertainty are highlighted based on entries in Table \ref{t-tlr}; 
the result of \cite{perrin-2004-01} is also overplotted.

\placefigure{Please place this on top of page7}

\begin{figure*}[t!]
\centering{
\includegraphics[width=\textwidth]{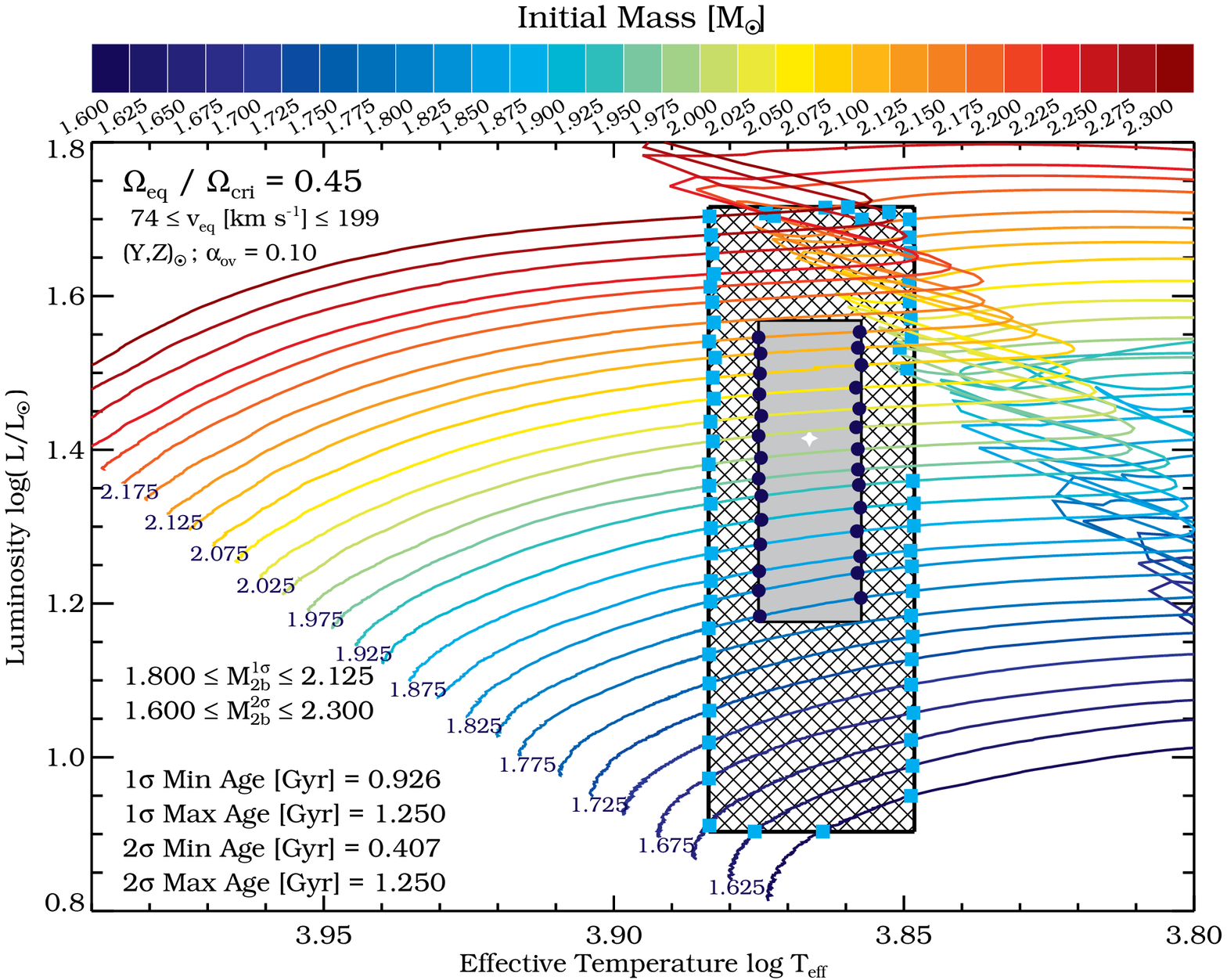}
\caption{A close zoom into Figure \ref{f-hrd-all} shows MESA tracks of $\alpha^2$ Her B within its observed 1$\sigma$ and 2$\sigma$ range of 
$\teff$ and $\log(L/L_\odot)$ (Table \ref{t-alf-sys}).
The age and mass of this star inferred from its position on HR diagram is listed in Table \ref{t-age} and Eq. \ref{e-M2b}, respectively.
The color coding is based on the initial mass for each track.}\label{f-alf-2-B} }
\end{figure*}

\subsection{Constraining the Age of the $\alpha$ Herculis System}\label{s-result}
From Figure \ref{f-hrd-all} and the measured physical properties of $\alpha^2$ Her B  (in Table \ref{t-alf-sys} and Section \ref{s-alf-her-sys})
this A9 IV-V star is either in the core hydrogen-burning phase, or has just entered the sub-giant phase.
Hence, it is the least evolved (and least massive) member of the system. 
Because the main sequence evolution of stars is understood with more certainty \citep[see][]{langer-2012-01},
the model inference for $\alpha^2$ Her B is more robust than for the other two components.
Consequently, we base the estimate of the \textit{age} of the system on the age we infer for the A9 IV-V star $\alpha^2$ Her B.
In other words, we assume that 
the only reason for the differences in evolutionary status of the three $\alpha$ Herculis stars lies in their differences in initial masses.

Figure \ref{f-alf-2-B} enlarges a small portion of the grid (i.e. Figure \ref{f-hrd-all}), 
and shows models with their respective luminosity and effective temperature
lying in the 1$\sigma$ (filled gray) and 2$\sigma$ (cross hatched) boxes of uncertainty for $\alpha^2$ Her B (see Table \ref{t-alf-sys}).
The filled symbols mark where the tracks enter/exit the highlighted zones, and where we measure the model ages.
The lower and upper age limits of the system are assessed based on the age at these flagged points.
We address these as the \textit{age constraints}.
Our results within the 1$\sigma$ and 2$\sigma$ uncertainties in $\teff$ and $\log(L/L_\odot)$ are summarised in Table \ref{t-age}.
The timesteps (around the flagged points in Figure \ref{f-alf-2-B}) are approximately two orders of magnitude smaller than the inferred ages.

  \begin{table}[h!]
  \begin{center}
  \caption{Model Age Estimation for $\alpha^2$ Her B.} \label{t-age} 
  \normalsize
  \begin{tabular}{lcc}
  \hline \hline
  Uncertainty & Minimum & Maximum  \\
   & [Gyr]        & [Gyr]         \\
  \hline 
  $1\sigma$ & 0.926 & 1.250 \\
  $2\sigma$ & 0.407 & 1.250 \\
  \hline
  \end{tabular}
  \end{center}
  \end{table}

We repeat the same procedure for M12 tracks.
For the age of $\alpha^2$ Her B, we find different results: with $1\sigma$ uncertainty, the age ranges from 0.787 to 1.452 Gyr, 
and similarly, with 2$\sigma$ uncertainty it ranges from 0.734 to 1.719 Gyr.
With respect to M12, the MESA ages roughly differ 14\% to 53\%.
We cannot extend this comparison to E12 and L12 tracks, as their coarse mass spacing does not allow such.
In Section \ref{ss-comp-age}, we address this age comparison again.

\subsection{Masses of $\alpha$ Herculis Stars from the HR Diagram}\label{ss-masses}
We designate the initial masses of $\alpha^1$ Her, $\alpha^2$ Her A and $\alpha^2$ Her B by
$M_{1}$, $M_{\rm 2a}$ and $M_{\rm 2b}$, respectively.
From the assumption that the differences in the current evolutionary status of $\alpha$ Herculis stars have their origins in their initial masses, 
$M_1$, $M_{\rm 2a}$ and $M_{\rm 2b}$ must obey this inequality
\begin{equation}\label{e-mass-inequality}
M_{\rm 2b} \, < \, M_{\rm 2a} \, < \, M_1.
\end{equation}

From Figure \ref{f-alf-2-B}, it is straightforward to find the most viable mass for $\alpha^2$ Her B
\begin{equation}\label{e-M2b}
\begin{array}{l}
1.800 \leq M_{\rm 2b}^{1\sigma}\, [M_\odot] \leq 2.125, \\
1.600 \leq M_{\rm 2b}^{2\sigma}\, [M_\odot] \leq 2.300.
\end{array}
\end{equation}
M12 tracks, strikingly, give $1.80 \leq M_{\rm 2b}^{1\sigma} [M_\odot]  \leq 2.10$ and $1.60 \leq M_{\rm 2b}^{2\sigma} [M_\odot] \leq 2.30$ 
in close agreement with Eq. \ref{e-M2b}.
Due to the coarse mass spacing in E12 and L12, we decide not to assess masses from their tracks.

\begin{figure*}[t!]
\begin{minipage}{0.49\linewidth}
\includegraphics[width=\columnwidth]{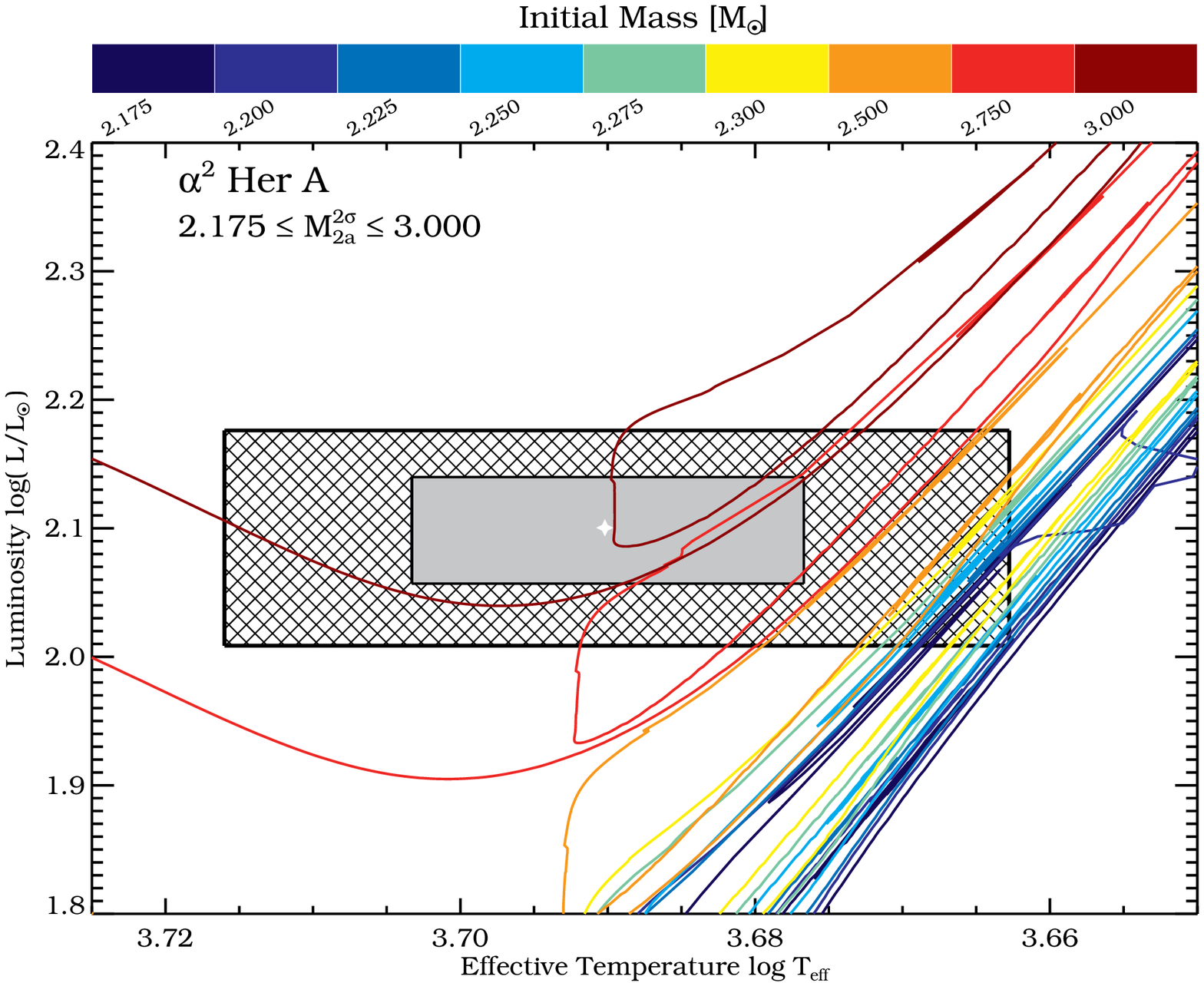}
\caption{MESA tracks of $\alpha^2$ Her A, within the 2$\sigma$ observed range of 
$\teff$, luminosity, and ages from Table \ref{t-age}. 
Eq. \ref{e-M2a} gives the possible mass of this star.
The color coding is based on the initial mass for each track.}\label{f-alf-2-A} 
\end{minipage}
\vspace*{0.02\linewidth}
\begin{minipage}{0.49\linewidth}
\includegraphics[width=\columnwidth]{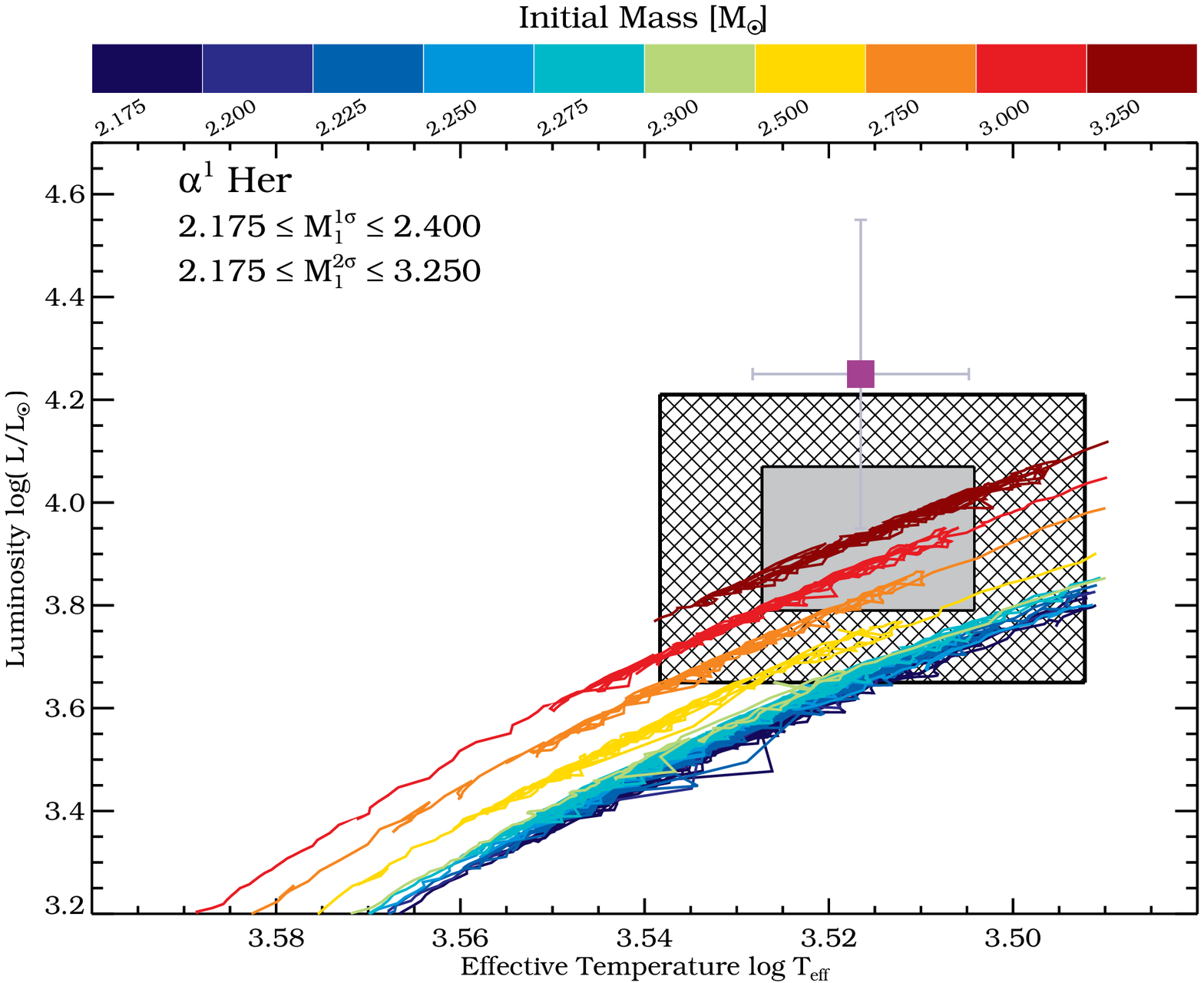} 
\caption{MESA tracks of $\alpha^1$ Her within the observed range of temperatures and luminosities (Table \ref{t-alf-sys}), 
and ages constrained from $\alpha^2$ Her B (Figure \ref{f-alf-2-B} and Table \ref{t-age}). 
The observation with errorbars is reproduced from \cite{perrin-2004-01}. 
The color coding is based on the initial mass for each track.}\label{f-alf-1} 
\end{minipage}
\end{figure*}

We permitted models with slightly higher masses than in Eq. \ref{e-M2b} to deplete their core helium content, to ascend the AGB, and to reach $\teff\leq3100$ K.
Our strategy is to tightly bind $M_1$ and $M_{\rm 2a}$ within 1$\sigma$ (and 2$\sigma$) uncertainties to find those tracks that 
\textit{simultaneously} match the observed $\teff$ and $\log L$ of these stars, in addition to their ages lying between the minimum and maximum age
of the system from Table \ref{t-age}.
Figure \ref{f-alf-2-A} shows $\alpha^2$ Her A on the HRD.
Only the tracks within 2$\sigma$ box can satisfy the above conditions;
therefore, the initial mass range for this star is
\begin{equation}\label{e-M2a}
\begin{array}{l}
2.175 \leq M_{\rm 2a}^{2\sigma}\, [M_\odot] \leq 3.000.
\end{array}
\end{equation}
where the uncertainty is not larger than 0.05 $M_\odot$.
One of the following evolutionary scenarios applies to $alpha^2$ Her A: it is ascending the RGB, has just ignited helium
in the core, or is on the early-AGB phase.

Figure \ref{f-alf-1} shows the expected location of the primary $\alpha^1$ Her on HRD based on the observations of \cite{perrin-2004-01} and the present work 
(Tables \ref{t-alf-sys} and \ref{t-tlr}). 
There is reasonable agreement of the luminosity of the primary star with the two approaches.
Similar to the previous stars, we assess the evolutionary initial mass of the primary based on its age and location on HRD as
\begin{equation}\label{e-M1}
\begin{array}{l}
2.175 \leq M_{\rm 1}^{1\sigma}\, [M_\odot] \leq 2.400, \\
2.175 \leq M_{\rm 1}^{2\sigma}\, [M_\odot] \leq 3.250.
\end{array}
\end{equation}
The solar-type pulsation pattern in the primary is already established \citep{bedding-2003-01,kiss-2006-01,moravveji-2010-01}. 
Based on this fact, we earlier estimated the mass of this star to be $2.5_{-1.1}^{+1.6}\, M_\odot$ \citep{moravveji-2011-01}
using the asteroseismic mass and radius scaling laws \citep{huber-2011-01}.
Though the size of uncertainties are large, the seismic mass is consistent with Eq. \ref{e-M1}.
In Sections \ref{ss-comp-age} and \ref{ss-comp-yields}, we attempt to examine Eq. \ref{e-M1} based on the surface abundances of the primary.

It is worthwhile to mention that the model radius for $\alpha^1$ Her based on our grid lies in the range $200\,\lesssim\, R/R_\odot \,\lesssim\, 314$.
This agrees better with our inferred radius in Table \ref{t-tlr} than with the near-IR interferometric estimate (Eq. \ref{e-alf-her-R}).
We admit that our treatments of the envelope convection and that of the extended atmosphere of AGB stars in our MESA models are simplistic.

\subsection{Agreement in AGB Age Assessment}\label{ss-comp-age} 
\begin{figure}[b]
\includegraphics[width = \columnwidth]{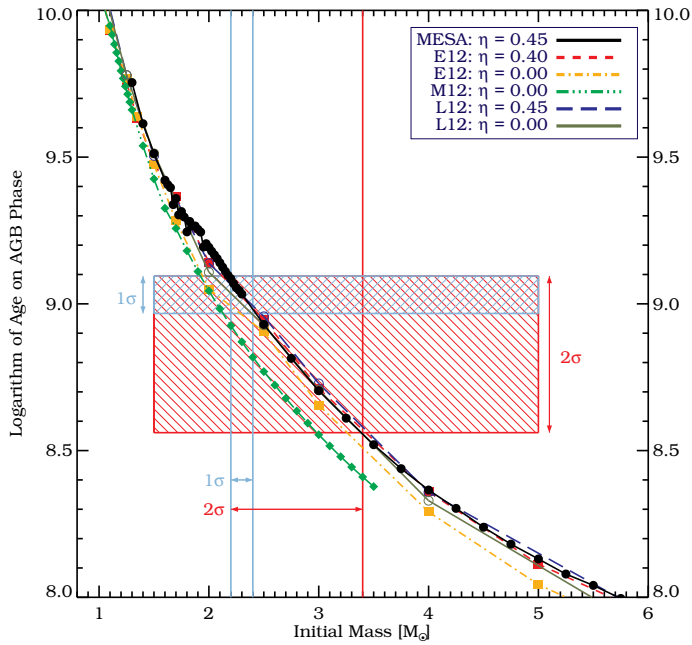}
\caption{Comparison of stellar age versus initial mass between MESA, E12, L12 and M12. 
The plotting symbols distinguish between different codes and their corresponding $\eta_{\rm rot}$.
The highlighted box marks the expected age of the $\alpha$ Her system from Table \ref{t-age}.}\label{f-comp-ages}
\end{figure}

Figure \ref{f-comp-ages} shows the final (i.e. AGB) lifetime of tracks from MESA, E12, L12 and M12 versus their corresponding initial masses.
The ages given by M12 (green line) are significantly less than the rest of models, as the tracks terminate on the subgiant phase.
On the higher mass regime, the lifetime of rotating star tracks in logarithmic scale is about 0.1 dex higher than the non-rotating star tracks.
This is explained by extra engulfment of hydrogen fuel by the rotating core during the MS phase.
Among evolution tracks including rotation, those of L12 have higher ages.

\placefigure{Please place this figure in page 10.}

The highlighted 1$\sigma$ and 2$\sigma$ boxes in Figure \ref{f-comp-ages} show the upper and lower bounds 
of the age of the $\alpha$ Herculis system from Table \ref{t-age}.
The vertical lines show the initial masses of stars which can reach the AGB phase within the given age (in agreement with Eqs. \ref{e-M1}).
Based on this, the age constraint from Table \ref{t-age} is robustly independent of the stellar evolution code used.
This places $\alpha^1$ Her among the few AGB stars in our galaxy with known ages.

\subsection{Surface Abundance Ratios of Carbon and Oxygen Isotopes}\label{ss-comp-yields}
The literature on the spectroscopic abundance analyses of $\alpha^1$ Her is, surprisingly, scarce.
\cite{harris-1984-01} measure the surface ratios of the key CNO processed species in the atmosphere of $\alpha^1$ Her.
They are $^{12}$C/$^{13}$C = $17\pm4$, $^{16}$O/$^{17}$O = $180_{-50}^{+70}$ and $^{16}$O/$^{18}$O = $550_{-175}^{+225}$.
The large uncertainty in the latter is not constraining, and we exclude it from our analysis.
To our knowledge, there is no record on the detection Li and/or Tc on $\alpha^1$ Her; 
we conservatively interpret this as the hot bottom burning not occurring in $\alpha^1$ Her,
and the mass being below nearly $\sim$ 4 $M_\odot$.
This complies with Eq. \ref{e-M1}.

\cite{el-eid-1994-01} employed these abundance ratios, and concluded that the mass of $\alpha^1$ Her lies in the interval 5 and 7 $M_\odot$.
However, the models calculated by El Eid did not include rotational and overshooting mixing.
During the past two decades, there have been major improvements in the input physics to the stellar evolution codes, 
mainly to the opacity, EOS and nuclear reaction rates.
For this reason, we repeat the same exercise as in \cite{el-eid-1994-01} with \texttt{MESA}. 

\begin{figure*}
\begin{minipage}{0.49\linewidth}
\includegraphics[width=\textwidth]{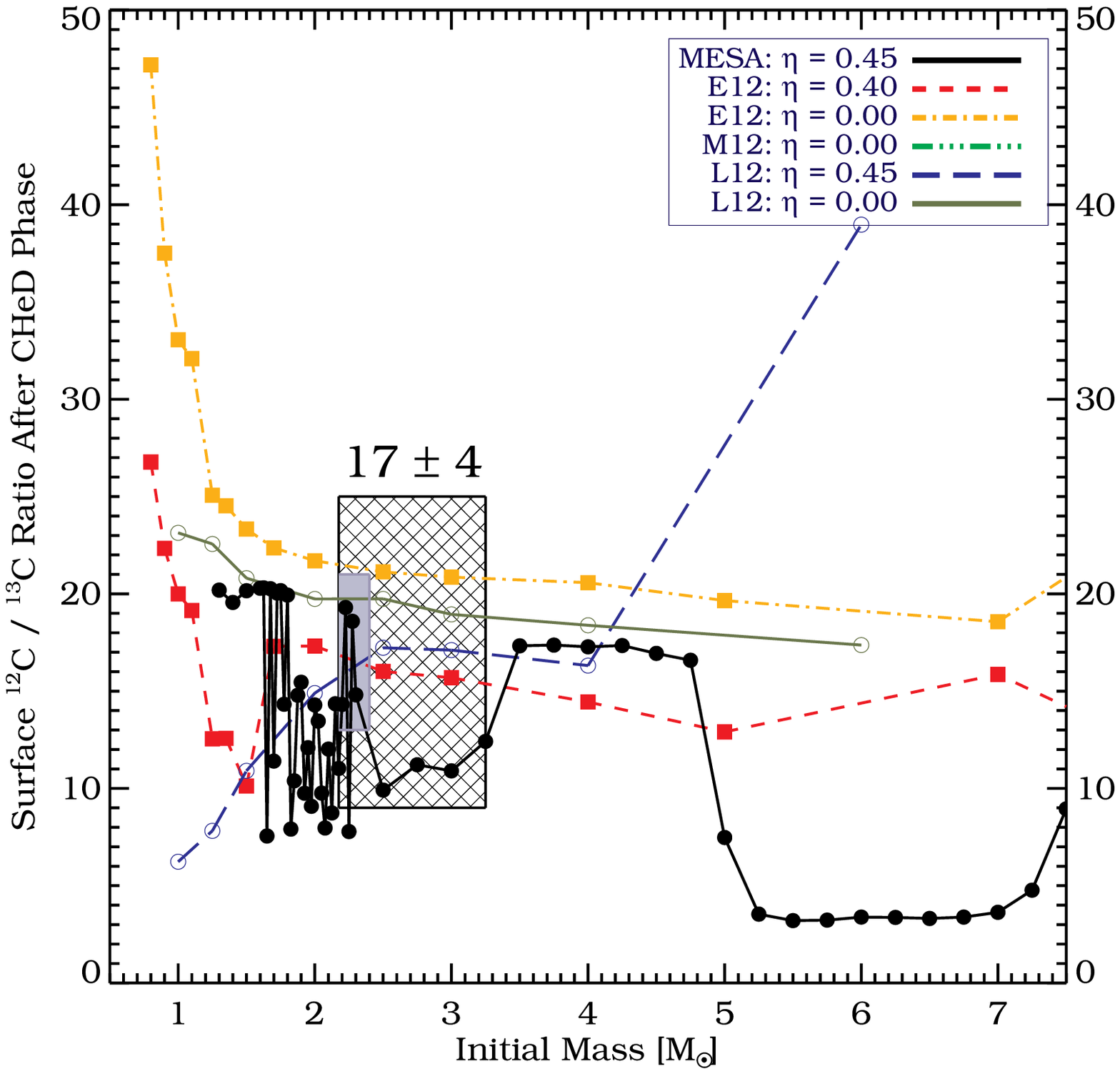}
\end{minipage}
\vspace*{0.02\linewidth}
\begin{minipage}{0.49\linewidth}
\includegraphics[width=\textwidth]{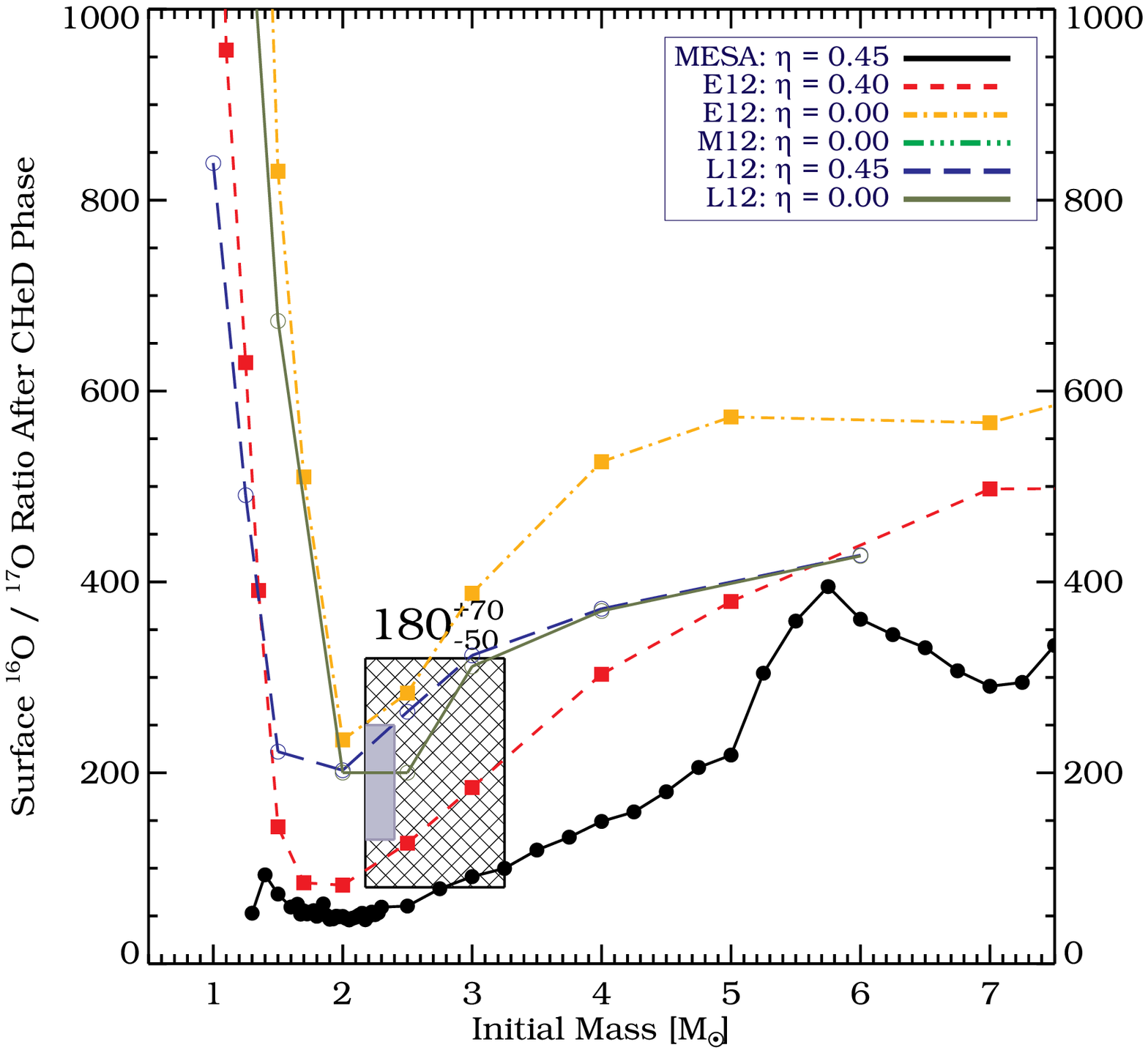}
\end{minipage}
\caption{Variation of the Surface abundance ratios of $^{12}$C/$^{13}$C (left) and $^{16}$O/$^{17}$O (right) with initial ZAMS masses for an M5 AGB star.
 We compare similar predictions from MESA, \cite[][E12]{ekstrom-2012-01} and \cite[][L12]{lagarde-2012-01}, and for different choices of
 the initial rotation rate $\eta_{\rm rot}$. 
 For $\alpha^1$ Her, the measured ratios within 1$\sigma$ and 2$\sigma$ boxes from \cite{harris-1984-01} are highlighted with 
 shaded and hatched strips, respectively. The mass intervals are adopted from Eqs. \ref{e-M1}.}\label{f-ratios}
\end{figure*}

Figure \ref{f-ratios} shows the surface abundance ratios of $^{12}$C/$^{13}$C (left panel) and $^{16}$O/$^{17}$O (right panel) versus the initial model masses.
We designate these by $r_1$ and $r_2$, respectively.
The observations from \cite{harris-1984-01} within the 1$\sigma$ and 2$\sigma$ uncertainties are highlighted.
These two ratios magnify the net contribution from convective and extra mixing mechanisms during the evolution history of the models.
We compare the same yields from MESA, E12 and L12 for their rotating and non-rotating tracks.
For the large departure of $v_{\rm eq}$ in L12 from MESA and E12, we subsequently present but do not discuss the surface abundance 
ratios from their rotating stellar tracks.

An inspection of $r_1=^{12}$C/$^{13}$C ratio (Figure \ref{f-ratios} left) indicates the different predictions made by different codes.
In E12, $r_1$ declines monotonically with the model mass (orange and red squares).
The inclusion of rotation gradually mixes extra $^{13}$C to the surface and $r_1$ is smaller for rotating tracks
compared to their non-rotating counterparts.
The non-rotating case of L12 (green empty circles) follows the same trend as E12.
In MESA (black filled circles), $r_1$ is irregular on the low-mass regime, and then exhibits a clear variable trend on the higher mass end.
Thus, it is not straightforward to assess the 1$\sigma$ and/or 2$\sigma$ mass of $\alpha^1$ Her with any certainty.

The $r_2=^{16}$O/$^{17}$O ratio (Figure \ref{f-ratios} right) shows nearly the same behaviour in all codes except MESA:
declining sharply with increasing model mass, reaching a dip around $\sim\,2M_\odot$, and rising again.
In E12, the gradual surface enrichment of $^{17}$O by rotation during the MS enforces a deeper dip.
The results of L12 are roughly between those of E12.
In MESA, the combined effects of atomic diffusion and rotational mixing result in the highest surface $^{17}$O enrichment which suppresses $r_2$.
For massive AGBs, the MESA predictions differ from the other models.
Once more, the mass assessment for $\alpha^1$ Her is not necessarily agreeing between different codes:
with the rotating E12 tracks, we find $1.4\lesssim M_1^{2\sigma}\lesssim4.2$,
with L12 we find $1.4\lesssim M_1^{2\sigma}\lesssim 3.2$,
and with MESA, we find $2.9\lesssim M_1^{2\sigma}\lesssim 5.3$.
We find none of the mass assessments in good agreement with the predictions of Eq. \ref{e-M1}.
Therefore, we do not succeed to fine-tune $M_1$ by using surface abundance ratios, $r_1$ and $r_2$.

\section{Discussion and Concluding Remarks}\label{s-conclusion}
In Sections \ref{s-calib} and \ref{s-lum-rad}, we propose a photometric method using Wing ABC filters to exploit the effective temperature (Eq. \ref{e-teff-calib}) and 
luminosity (Eqs. \ref{e-bol-corr-c} to \ref{e-bol-lum}) of evolved - mid-K to mid-M spectral type - stars in agreement with near infrared interferometry.
On one hand, direct measurement of angular diameter and $\teff$ for stars based on long-baseline interferometry has some shortcomings:
\begin{enumerate}[(i)]
\item there are currently few actively operating interferometers that are accessible for the broad astronomy community,
\item a limited number of stars fall within the observability of current instruments, according to their apparent magnitude and apparent angular diameter.
\end{enumerate}
On the other hand, our proposed small-aperture photometry does not suffer these limitations, and can be applied to individual evolved stars.

To deduce the mass and age of $\alpha$ Herculis stars, we used a grid of stellar evolutionary tracks.
The assumptions, simplifications and uncertainties in the physical parameters of the model translates into significant uncertainties in calculating model masses, radii and ages.
\citet{basu-2012-01} provide an in-depth analysis of these grid-based approaches. 
They estimate that the accuracy of mass evaluation without inclusion of additional seismic information is at least 8\%.
We conclude from Figure \ref{f-comp-ages} that the model ages and masses calculated by MESA, E12, L12 and M12 are in satisfactory agreement.
This is not a surprise as far as the four codes we are comparing employ very similar nuclear reaction rates\footnote{
All employed codes in this study use NACRE \citep{angulo-1999-01} thermonuclear reaction rates with updates to 
$^{14}$N(p,$\gamma$)$^{16}$O, triple-$\alpha$,
$^{14}$N($\alpha,\gamma$)$^{18}$F and $^{12}$C($\alpha,\gamma$)$^{16}$O reactions.}.
This result supports the stringency of model-dependent age determination approaches, such as asteroseismology of red giants in clusters 
\citep{basu-2011-01, miglio-2012-01}.

\begin{figure}[b]
\centering{
\includegraphics[width=\columnwidth]{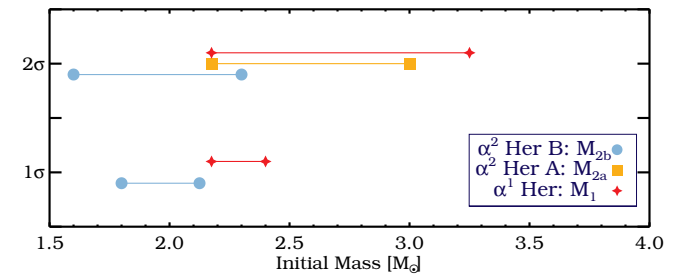}
\caption{Distribution of stellar mass in the $\alpha$ Herculis system.
See the text and Eqs. \ref{e-M2b} to \ref{e-M1} for explanations.
Yet, we cannot apply Eq. \ref{e-mass-inequality} to set more strict limits on the mass of each star in the system.
The mass assessment from Figure \ref{f-ratios} is non-constraining.}\label{f-comp-mass}
}
\end{figure}

Figure \ref{f-comp-mass} summarizes our results on the mass distribution in the $\alpha$ Herculis system within 1$\sigma$ and 2$\sigma$ uncertainties.
It is a collection of the results from Eqs. \ref{e-M2b} to \ref{e-M1} which employ the position of $\alpha$ Herculis stars on the HRD 
(from Figures \ref{f-alf-2-B} to \ref{f-alf-1}) and the age constraint (Table \ref{t-age}).
From the condition that the masses of $\alpha$ Herculis stars must not overlap (Eq. \ref{e-mass-inequality}), 
we have no additional information that would limit the mass ranges for components of $\alpha$ Herculis system.

Figure \ref{f-ratios} shows very different patterns for $^{12}$C/$^{13}$C and $^{16}$O/$^{17}$O atmospheric abundance ratios.
It is difficult to judge whether the observed differences between the $^{12}$C/$^{13}$C and $^{16}$O/$^{17}$O trends in different codes is more of a 
physical nature or of computational origin;
this is beyond the scope of this study.
Yet, it calls for an observational calibration of surface abundances versus the global stellar parameters, such as mass and $\log g$,
which in return requires a high precision mass assessment.
The asteroseismology of oscillating red giants comes to the rescue, as it can indirectly provide measures of $\log g$ from seismic scaling laws as precise as 
$\sim1\%$ \citep[see e.g.][]{basu-2011-01}. 
Also see \cite{morel-2012-01} and \cite{thygesen-2012-01}. 
Once this is relaxed, surface abundance ratios might serve as an alternative technique in estimating the masses of stars
when seismic and/or binarity information is missing.


\section{Summary}\label{s-summary}
We carried out more than two decades of multicolor photometry of the nearby triple-stellar system, $\alpha$ Herculis, 
and devise a method to extract effective temperature and the bolometric luminosity of the primary star. 
For this, we use Wing ABC filters.
For $\alpha^1$ Her, we find $\teff=3280\pm87$ and $\log(L/L_\odot)=3.92\pm0.14$.
These agree with the near-infrared interferometric observations of \cite{perrin-2004-01} within the error bars.

We calculated a grid of 55 evolutionary tracks with MESA which incorporate the effects of stellar rotation.
The grid has the solar composition, and is calculated for the mass range 1.30 $M_\odot$ to 8.0 $M_\odot$.
Within 2$\sigma$ uncertainty, the $\alpha$ Herculis system has an age of 0.41 to 1.25 Gyr.
The inferred model age from MESA agrees with E12 and L12 tracks.
We consequently find that the initial masses of the stars in $\alpha$ Herculis system are distributed between 1.60 to 3.40 $M_\odot$, 
with the primary M5 Ib-II AGB star having the mass $2.175\leq M_1\leq 3.250$.
This result was independently reproduced by \cite{moravveji-2011-01}, by extending the seismic  scaling relations for RGBs to AGB stars.
This now settles the debates on the mass of $\alpha^1$ Her indicating a smaller value than formerly thought, and rejects its evolutionary status 
being a more massive red supergiant (like $\alpha$ Ori and $\alpha$ Sco; see Section \ref{s-mass-debate}).

Soon, Gaia will provide precise parallaxes for nearly half a million Galactic stars.
Multicolor photometry of M-type giants and supergiants in Wing ABC filters, when combined with such precise parallaxes, 
can provide the stellar effective temperatures and luminosities
at a precision comparable to or even better than the infrared interferometry.


\acknowledgments 
\textbf{Acknowledgments} 
We appreciate the comments from the anonymous referee that helped us improve this manuscript.
We thank Andrej Pr\v{s}a for granting us an access to the Villanova University computing facility, and Thomas Lebzelter and Achim Weiss for reading and commenting on this document. 
E.M. is grateful to Bill Paxton and Falk Herwig for many fruitful discussions about MESA, and also to the board of MESA for freely publishing the code.
We acknowledge using the Coyote IDL graphics packages made freely available by David Fanning.
The research leading to these results has received funding partly from the European Research Council under the European Community's Seventh Framework Programme
(FP7/2007--2013)/ERC grant agreement n$^\circ$227224 (PROSPERITY), and partly by the NSF/RUI grant AST-1009903 to Villanova University that we gratefully acknowledge.
\bibliographystyle{apj}
\bibliography{apj-jour,/Users/ehsan/my/papers/my-bib}
\appendix

\section{Wing C-Filter Bolometric Correction BC$_{\rm C}$}\label{app-abm}
The C-filter of Wing's 3-color system is centred in a continuum region free from strong absorption lines (see Fig \ref{f-spec} and Table \ref{t-obs}).  
The central wavelength is at 1040 nm with a FWHM of 42 nm \cite{white-1978-01}.  
The transmitted flux through the filter measures near-IR apparent magnitudes that approximate bolometric magnitudes, as seen in Mira-variable light
curves near their energy maxima \citep{wing-1992-01}. 
Bolometric corrections BC$_{\rm C}$ between 1040 nm magnitudes and the UBV-based apparent bolometric magnitudes are computed for eight 
M4.8 to M5.1 calibration stars via Eq. \ref{e-app-1}.
\begin{equation}\label{e-app-1}
\mbox{BC}_{\rm C} = m_{\rm bol} - m_{\rm 1040}, 
\end{equation}
where $m_{\rm 1040}$ is the 1040 nm magnitude and is taken from \cite{wing-1978-01} for each star.
The UBV-based apparent bolometric magnitude, $m_{\rm bol}$, is calculated from Eq. \ref{e-app-2}
\begin{equation}\label{e-app-2}
m_{\rm bol} = V + \mbox{BC}_{\rm V},
\end{equation}
The V-band magnitudes are taken from the Bright Star Catalog \citep{hoffleit-1982-01} or the Simbad Astronomical Database.
Using the bolometric corrections in Table 5 of \cite{levesque-2005-01}, a second-order polynomial is generated to calculate unique
bolometric corrections that are dependent on the spectral sub-types of the eight calibration stars. 
This second-order polynomial is given in Eq. \ref{e-app-3}
\begin{equation}\label{e-app-3}
\mbox{BC}_{\rm V} = -0.0282\,x^2 - 0.039\,x - 1.1703, \quad \chi^2_{\rm red} = 0.9821.
\end{equation}
where $x$ represents the numerical part of the spectral sub-type plus one.  
For example, to compute the BC$_{\rm V}$ of a M4.9 star, $x$ = 5.9 in Eq. \ref{e-app-3}. 
Table \ref{t-bol-corr} lists the calibration stars with their spectral types, V, BC$_{\rm V}$, $m_{\rm 1040}$, and BC$_{\rm C}$ magnitudes, respectively.  
An average is then taken of all eight bolometric corrections to C to yield the final correction value itself.  
The bolometric correction to C for each star is given in the last column in Table \ref{t-bol-corr}, and the final bolometric correction to C is BC$_{\rm C}=1.735\pm0.030$.
This value is added to the color-corrected C-filter 1040 nm bolometric magnitudes, and these final resulting magnitudes are then used to compute the luminosities 
in Figure \ref{f-tlr} and Table \ref{t-tlr}.

\begin{deluxetable}{lcccccccc}
 \tablecaption{\cite{wing-1978-01} calibration stars used to compute Wing C-filter magnitude corrections.
 Column references are included. \label{t-bol-corr} }
 \tablecolumns{9}
  \tablehead{  \colhead{HR} & \colhead{Spectral} & \colhead{$V$} & \colhead{} & \colhead{BC$_{\rm V}$} & \colhead{}  & \colhead{Wing 1040-nm} & \colhead{} & \colhead{BC$_{\rm C}$}   \\
  \colhead{Number} & \colhead{Type} & \colhead{[mag]} & \colhead{} & \colhead{[mag]} & \colhead{} & \colhead{[mag]} & \colhead{} & \colhead{[mag]} \\
   }  %
  \startdata
     (a)   &     (b)     & (a)    &         & (Eq. \ref{e-app-3}) & & (b) & & (Eq. \ref{e-app-1}) \\
     \tableline
     85   &     M4.8 & 5.12 & $+$ & ($-$2.345) & $-$ & 0.97 & $=$ & 1.805 \\
   1722 &     M4.8 & 5.65 & $+$ & ($-$2.345) & $-$ & 1.71 & $=$ & 1.595 \\
   4949 &     M4.8 & 5.66 & $+$ & ($-$2.345) & $-$ & 1.53 & $=$ & 1.785 \\
   7804 &     M4.8 & 5.55 & $+$ & ($-$2.345) & $-$ & 1.60  & $=$ & 1.605 \\
   4045 &     M4.9 & 6.30 & $+$ & ($-$2.382) & $-$ & 2.15 & $=$ & 1.768 \\
   5192 &     M5.0 & 4.19 & $+$ & ($-$2.419) & $-$ &(-0.03)& $=$& 1.801 \\
    587  &     M5.1 & 5.51 & $+$ & ($-$2.457) & $-$ & 1.31 & $=$ & 1.743 \\
   4909 &     M5.1 & 5.84 & $+$ & ($-$2.457) & $-$ & 1.60 & $=$ & 1.783 \\
   \tableline
    Ave. Spec. & M4.9   &       &             &     &        & Average & $=$ & 1.735 \\
                    &            &       &             &     &        & Std. Dev. & $=$ & 0.085 \\
                    &            &       &             &     &        & Std. Err.  & $=$ & 0.030 \\
	\enddata
	\tablerefs{ (a) \cite{hoffleit-1982-01}, (b) \cite{wing-1978-01}. }
  \end{deluxetable}

\section{Light Curves in Johnson VRI and Wing ABC Filters}
\begin{figure}[t]
\centering{
\includegraphics[width=\textwidth]{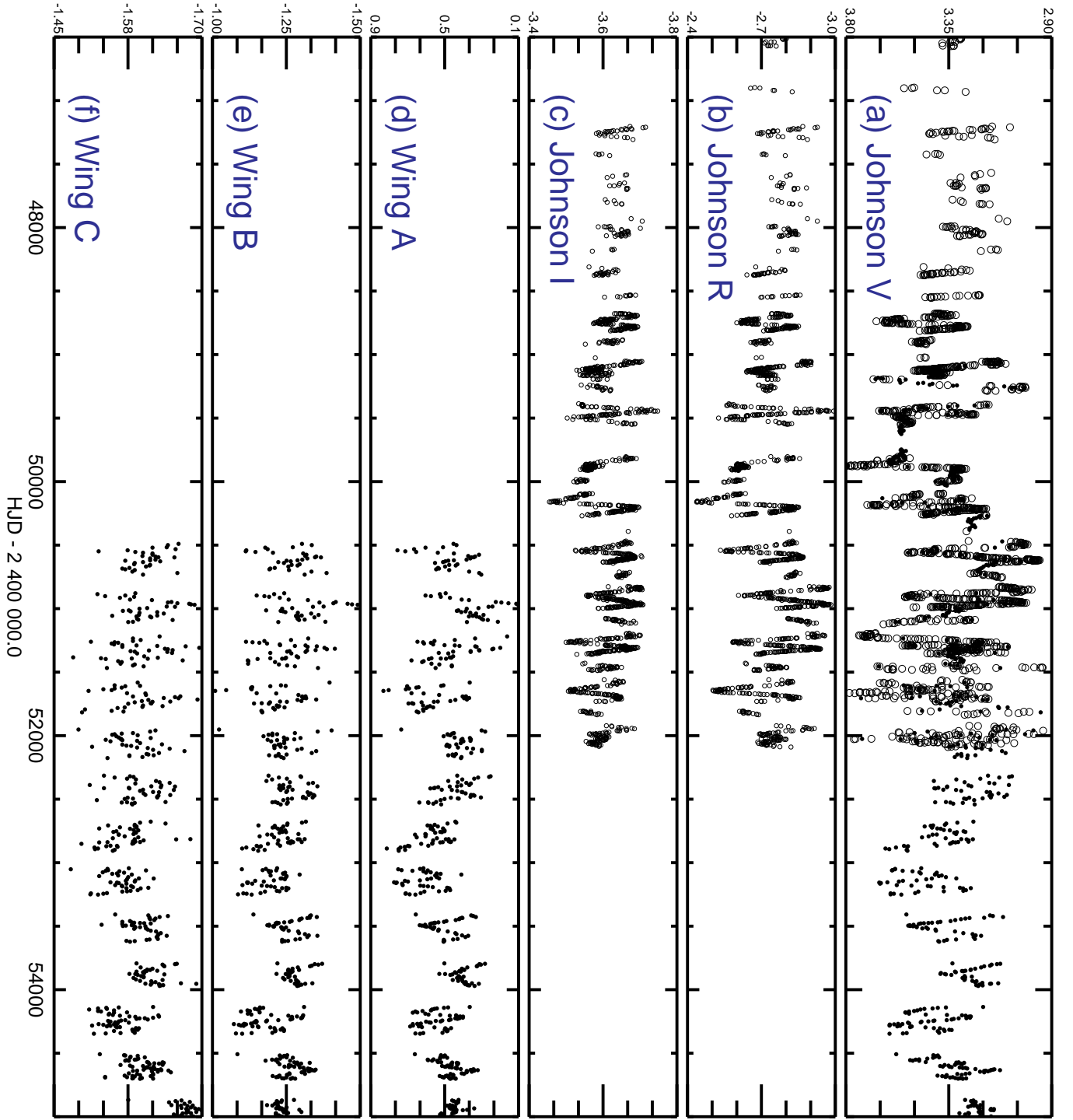}
\caption{
Multi-color, multi-epoch (23-years) photometry of the $\alpha$ Herculis system. 
Light curves are presented in increasing central filter wavelengths, and are collected with the Johnson VRI and the Wing ABC filters.
Empty circles ($\circ$) are observations collected at the TSU and filled circles ($\bullet$) are those collected at the VU.
For more details of the dataset, see Section \ref{s-phot} and Table \ref{t-obs}.
In panel (a), the overlap between the TSU and VU observations are in excellent agreement, and fill out one another's gaps. }\label{f-obs-all} }
\end{figure}

\end{document}